\documentclass[aps,pre,preprint,groupedaddress,showpacs]{revtex4}

\usepackage{graphicx,amssymb,amsmath}

\newcommand{\pder}[2]{\frac{\partial #1}{\partial #2}}
\newcommand{\rmd}{{\rm d}}
\newcommand{\rme}{{\rm e}}
\newcommand{\deriv}[2]{\frac{{\rmd} #1}{{\rmd} #2}}
\newcommand{\tsfrac}[2]{{\textstyle\frac{#1}{#2}}}

\newcommand{\be}{\begin{equation}}
\newcommand{\ee}{\end{equation}}
\newcommand{\beq}{\begin{equation}}
\newcommand{\eeq}{\end{equation}}
\newcommand{\bea}{\begin{eqnarray}}
\newcommand{\eea}{\end{eqnarray}}
\newcommand{\bean}{\begin{eqnarray*}}
\newcommand{\eean}{\end{eqnarray*}}
\newcommand{\rmi}{{\rm i}}
\newcommand{\ep}{\epsilon}
\newcommand{\cc}{{\mbox{c.c.}}}

\begin{document}

\title{The Nikolaevskiy equation with dispersion}


\author{Eman Simbawa}
\email[]{pmxes3@nottingham.ac.uk}
\author{Paul C. Matthews}
\email[]{paul.matthews@nottingham.ac.uk}
\author{Stephen M. Cox}
\email[]{stephen.cox@nottingham.ac.uk}
\affiliation{School of Mathematical Sciences, University of Nottingham,
Nottingham NG7 2RD, United Kingdom}


\date{\today}

\begin{abstract}

The Nikolaevskiy equation was originally proposed as a model for
seismic waves and is also a model for a wide variety of systems
incorporating a neutral ``Goldstone'' mode, including
electroconvection and reaction--diffusion systems.
It is known to exhibit chaotic dynamics at the onset of
pattern formation, at least when the dispersive terms in the equation are
suppressed, as is commonly the practice in previous analyses. In this
paper, the effects of reinstating the dispersive terms are examined. It is
shown that such terms can stabilise some of the spatially periodic traveling
waves; this allows us to study the loss of stability and transition to
chaos of the waves. The secondary stability diagram (``Busse
balloon'') for the traveling waves can be remarkably complicated.

\end{abstract}


\maketitle

\section{Introduction}\label{sec:intro}

In 1989, Nikolaevskiy~\cite{Nik89} derived a model for longitudinal
seismic waves, in the form of a one-dimensional partial differential
equation for a displacement velocity. Although Nikolaevskiy's equation
included dispersive terms, most subsequent analysis has treated a
simplified version of the PDE, in which these terms are omitted. This
reduced form is now generally known as the {\em Nikolaevskiy equation},
which may be written in the form
 \begin{equation}
\pder{u}{t}=-\pder{^2}{x^2}\left[r-\left(1+\pder{^2}{x^2}\right)^2\right]u
-u\pder{u}{x},
\label{eq:nik0}
 \end{equation}
where $r$ is a control parameter. The equation (\ref{eq:nik0}) has
been proposed as a model for several other physical systems, including
phase instabilities in reaction--diffusion equations~\cite{FujiYam},
electroconvection~\cite{TribVel} and transverse instabilities of
fronts~\cite{CMdamp}. More generally, (\ref{eq:nik0}) can be regarded
as a simple model of a pattern-forming system with an instability at
finite wavenumber and a neutral ``Goldstone'' mode arising from
symmetry~\cite{MattCox00,TribVel}.

The uniform state $u\equiv0$ of (\ref{eq:nik0}) becomes
unstable at $r=0$ to spatially periodic ``roll'' solutions, with
wavenumbers around $k=1$. However, these, in turn, are themselves all
unstable at onset in sufficiently large domains~\cite{TribVel}; this
unusual instability arises from the neutral mode at wavenumber $k=0$.
In fact, numerical simulations show that the Nikolaevskiy 
equation exhibits spatiotemporal chaos at onset~\cite{TribTsu,MattCox00}. The
scalings associated with this chaotic regime are unusual in pattern
forming systems~\cite{MattCox00,SakTan}, and this interesting feature of the
equation has stimulated significant investigation~\cite{WittPoon}.

Although in some applications (such as the instability of 
fronts~\cite{CMdamp}) the omission of dispersive terms is justified on 
symmetry grounds, this is not the case in the original 
context of a model for seismic waves~\cite{Nik89}. 

Earlier work that {\em has} considered the effects of dispersion includes 
the paper of Malomed~\cite{MalZero}, who reinstated one dispersive term in 
the Nikolaevskiy equation and analysed the secondary stability of 
traveling-wave solutions by means of coupled Ginzburg--Landau-type 
equations for the amplitude of the traveling waves and a large-scale 
mode. His results showed that dispersion could stabilize waves; however, 
his derivation was not entirely asymptotically 
self-consistent~\cite{TribVel}. Kudryashov and Migita~\cite{KudMig} 
showed, on the basis of numerical simulations, that traveling waves can 
be stabilized by the presence of dispersive terms in the Nikolaevskiy 
equation. It is also known that in the related Kuramoto--Sivashinsky 
equation, the introduction of a dispersive term can stabilize periodic 
traveling waves~\cite{Kawa}.

Our aim in this paper is to provide
a systematic examination of the effects of dispersion.
By varying the parameters corresponding to dispersion,
we can find when dispersion stabilizes traveling waves
and investigate how the chaotic state in the non-dispersive equation
arises as the dispersion is reduced.

In the following section we give the form of the equation and the 
traveling waves under consideration. Computational results on the
stability of these waves are given in Sec.~III. 
The stability analysis of the
waves is complicated and depends on the magnitude of the dispersion 
terms; three different scalings are considered in Secs.~IV, V and~VI. 
Sec.~VII illustrates some numerical simulations of 
the Nikolaevskiy equation with dispersion, and our conclusions are summarized 
in Sec.~VIII.

\section{The Nikolaevskiy equation with dispersion}\label{sec:nik}

We examine the Nikolaevskiy equation with dispersion in the form
 \begin{equation}
\pder{u}{t}=-\pder{^2}{x^2}\left[r-\left(1+\pder{^2}{x^2}\right)^2\right]u
-u\pder{u}{x}+\alpha\pder{^3u}{x^3}+\beta\pder{^5u}{x^5},
\label{eq:nikd}
 \end{equation}
where $\alpha$ and $\beta$ are the dispersion coefficients. This equation
is thus the one originally proposed by Nikolaevskiy~\cite{Nik89} (and
later examined in~\cite{KudMig,MalZero}), with all spatial derivatives up
to the sixth appearing on the right-hand side. In the numerical
simulations presented in Sec.~\ref{sec:numsim}, we shall impose the
periodic boundary condition
 \beq
u(x+D,t)=u(x,t)
\label{eq:bcs}
 \eeq
for some domain length $D$.

Before proceeding, we note that \eqref{eq:nikd} has the same Galilean
symmetry ($x\mapsto x+Vt$, $u\mapsto u+V$) as the nondispersive equation
\eqref{eq:nik0}. Moreover, in view of the Galilean symmetry and the
observation that, when the boundary condition \eqref{eq:bcs} is imposed,
 \[
\deriv{}{t}\int_0^D u(x,t)\,\rmd x=0,
 \]
the spatial average of $u$ may be set as zero (by transforming to a moving 
frame of reference if necessary). The reflection symmetry ($x\mapsto-x$, 
$u\mapsto-u$) of \eqref{eq:nik0} is broken by the presence of the 
dispersive terms. However, there is a symmetry $x\mapsto-x$, $u\mapsto-u$, 
$\alpha\mapsto-\alpha$, $\beta\mapsto-\beta$; as a consequence of this 
symmetry we need consider only the case $\beta \geq 0$.

\begin{figure}
\includegraphics[width=0.6\linewidth]{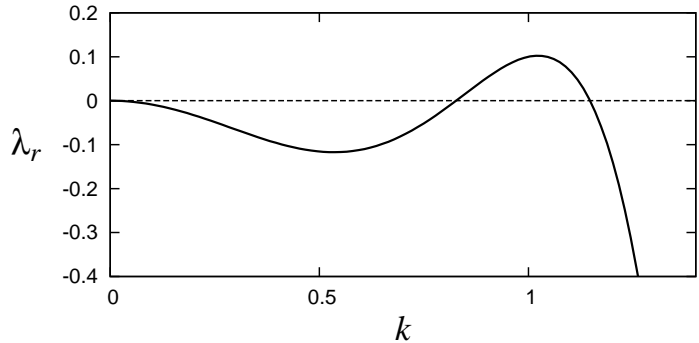}
\caption{Plot of $\lambda_r$ for the case $r=0.1$: note the linearly
growing modes with wavenumber around $k_c=1$, and the weakly damped
large-scale modes close to $k=0$.}

\label{fig:disprel}
\end{figure}

Linearization around the steady state $u\equiv0$ yields the dispersion
relation
 \[
\lambda=k^2\left[r-(k^2-1)^2\right]+\rmi k^3(k^2\beta-\alpha)
 \]
for Fourier modes proportional to $\rme^{\rmi kx+\lambda t}$. Thus in
general these perturbations take the form of traveling waves, with phase
speed
 \beq
c_p=-\frac{\lambda_i}{k}=k^2(\alpha-k^2\beta)
\label{eq:phase}
 \eeq
and group velocity
\beq
c_g=-\frac{\partial \lambda_i}{\partial k} = k^2(3\alpha-5\beta k^2).
\label{eq:group}
\eeq
The real part of the growth rate, $\lambda_r$, is plotted in
Fig.~\ref{fig:disprel}, for $r$ just above the threshold value $r_c=0$
for the onset of instability. This figure shows that there exists a band
around the critical wavenumber $k_c=1$ of linearly growing modes, and a
neutral mode at $k=0$ (the so-called ``Goldstone mode''), which
significantly affects the nonlinear dynamics of \eqref{eq:nikd}.

Just beyond the onset of instability of the zero solution, it is
straightforward to carry out a weakly nonlinear analysis of
\eqref{eq:nikd}, with
 \beq
r=\ep^2 r_2.
\label{eq:r2}
 \eeq
This analysis reveals that there are traveling-wave solutions of the form
 \beq
u\sim\ep a_0\rme^{\rmi k(x-st)}+\cc,
 \label{eq:TW}
 \eeq
where the wavenumber $k=1+\ep q$. The amplitude turns out to be given by
 \beq
a_0=6(r_2-4q^2)^{1/2}(1+\tsfrac{1}{36}(\alpha-5\beta)^2)^{1/2}
 \label{eq:TWamp}
 \eeq
and the speed of the wave is
 \beq
s=c_p-\tsfrac16\ep^2(r_2-4q^2)(\alpha-5\beta)+o(\ep^2),
 \label{eq:TWspeed}
 \eeq
where $c_p$ is given by \eqref{eq:phase} and the second contribution to $s$
reflects (weakly) nonlinear effects. So, regardless of the values of the
dispersion parameters $\alpha$ and $\beta$, such spatially periodic
solutions exist for $r_2>4q^2$. We now turn to the question of the
secondary stability of these solutions.

\section{Secondary stability of traveling waves: numerical results}
\label{sec:secstab}

In this section we first outline a numerical method for the 
calculation of the nonlinear traveling waves and their secondary
stability, and then give the results of these computations, showing
the stability boundaries of traveling
waves. 

\subsection{Numerical method for calculating secondary stability}

To calculate the secondary stability of a traveling wave solution for
given values of the parameters, we first find the traveling wave solution
$\bar u(x,t)=f(z)$, where $z=x-ct$. Here, $c$ is the nonlinear wave
speed, which in general is not exactly equal to the linear wave speed $c_p$
(\ref{eq:phase}). We approximate the solution
numerically using the truncated Fourier series
 \[
f(z)=\sum_{-N/2+1}^{N/2} \bar{u}_n\rme^{\rmi nkz}.
 \]
Substitution in \eqref{eq:nikd} (and calculation of the nonlinear term
pseudospectrally) yields a system of nonlinear equations (solved in
Matlab) for the Fourier coefficients of $f(z)$, together with $c$, which
is determined from
 \begin{equation}
c\int_{0}^{D}(f')^2\,\rmd z=
\alpha\int_{0}^{D}(f'')^2\,\rmd z -
\beta \int_{0}^{D}(f''')^2\,\rmd z
+\int_{0}^{D}f(f')^2\, \rmd z,
\label{eq:cphase}
 \end{equation}
where $D=2\pi/k$ is the length of the domain and $k$ is the wavenumber of
the solution under consideration. The expression \eqref{eq:cphase} follows
from multiplying \eqref{eq:nikd} by $f'(z)$ and integrating over the
domain, using integration by parts multiple times. To compensate for the
additional unknown $c$, we have an additional equation from the fact that
we may choose the phase of the wave, for example by specifying that
$\bar{u}_1$ is real.

After calculating the solution, we construct the eigenvalue problem
for perturbations. If we suppose that $u(x,t)=f(z)+\tilde u(x,t)$, then
substitution in \eqref{eq:nikd} yields the linearized perturbation
equation
 \begin{equation}
\pder{\tilde u}{t}=-\pder{^2}{x^2}\left[r-\left(1+\pder{^2}{x^2}\right)^2
\right]\tilde u
+\alpha\pder{^3{\tilde
u}}{x^3}+\beta\pder{^5{\tilde u}}{x^5}-f(z)\pder{\tilde u}{x}-
\tilde uf'(z).
\label{eq:31}
 \end{equation}
We take
 \[
\tilde u=\rme^{\sigma t+\rmi pz}\sum_{-N/2+1}^{N/2} v_n\rme^{\rmi nkz},
 \]
where all possible eigenfunctions may be captured by limiting
consideration to $-k/2\leq p\leq k/2$. The resulting eigenvalue equations
to determine the growth rate $\sigma$ are then
 \[
(\sigma-\rmi cK_n)v_n={\mathcal L} v_n-\sum_{-N/2+1}^{N/2}\rmi
K_mv_m\bar{u}_{n-m}-\sum_{-N/2+1}^{N/2}\rmi mkv_{n-m}\bar{u}_m,
 \]
where ${\mathcal L}=K_n^2(r-(1-K_n^2)^2)-\rmi\alpha K_n^3 +\rmi\beta
K_n^5$ and $K_n=p+nk$. The eigenvalues of this system are computed
numerically. By examining the largest real part of all eigenvalues
$\sigma$ for a large sample of values of $p$ in the relevant interval, we
determine whether the original traveling waves are stable or unstable. In
the following section we provide some stability diagrams based on the
above method.

In determining our results, we have been careful to check that: adequate
samples in $p$ are taken (too few, particularly for small values of $p$,
can lead one to miss certain small regions of instability); adequate Fourier
modes are taken in determining both the original solution and the
perturbations; adequate samples are taken in parameter space to determine
all regions of stable rolls. Typically, 300 values of $p$ are used,
with $N=16$.

\subsection{Results}

\begin{figure}

(a)\includegraphics[height=0.2\linewidth]{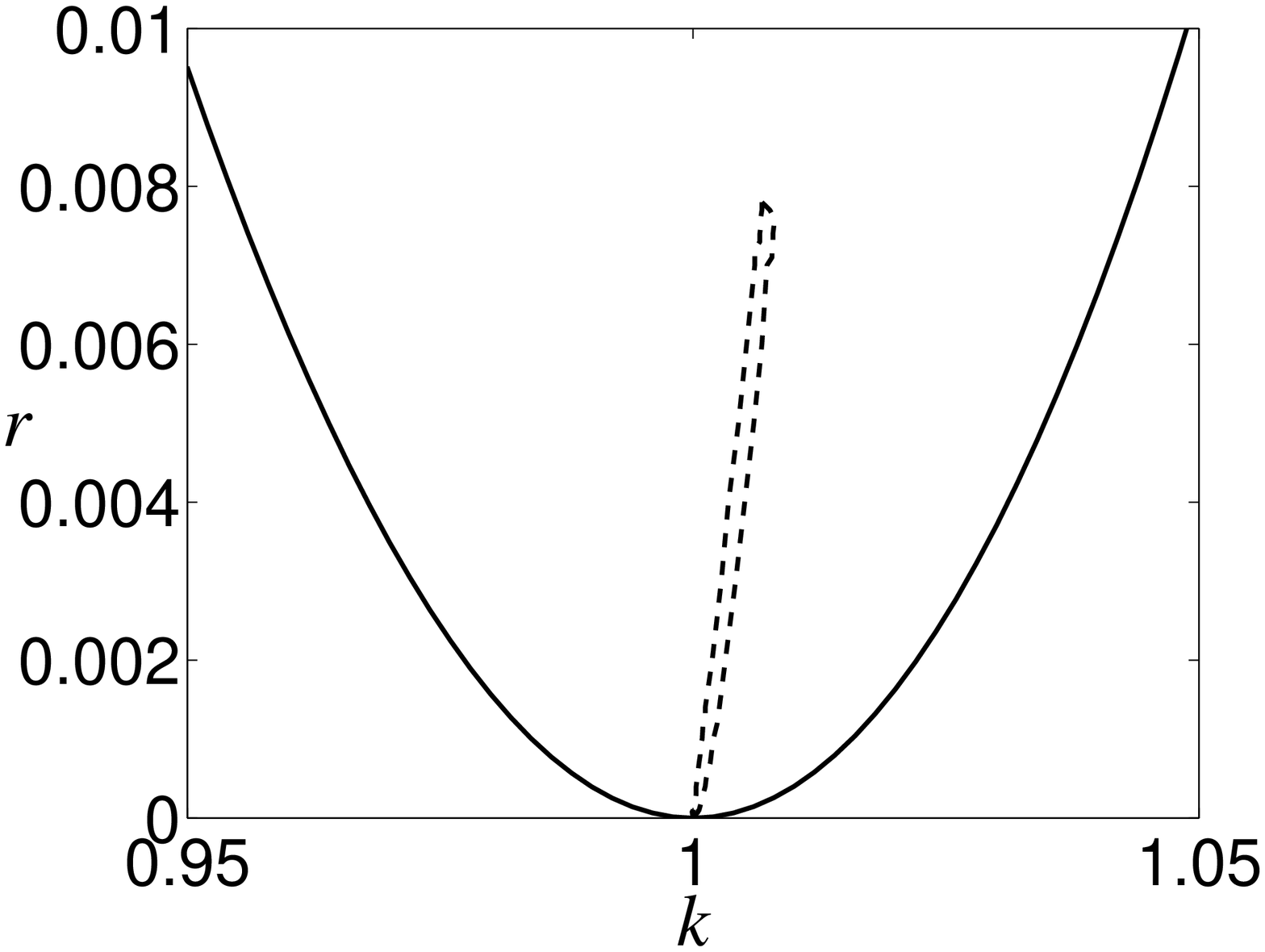}
(b)\includegraphics[height=0.2\linewidth]{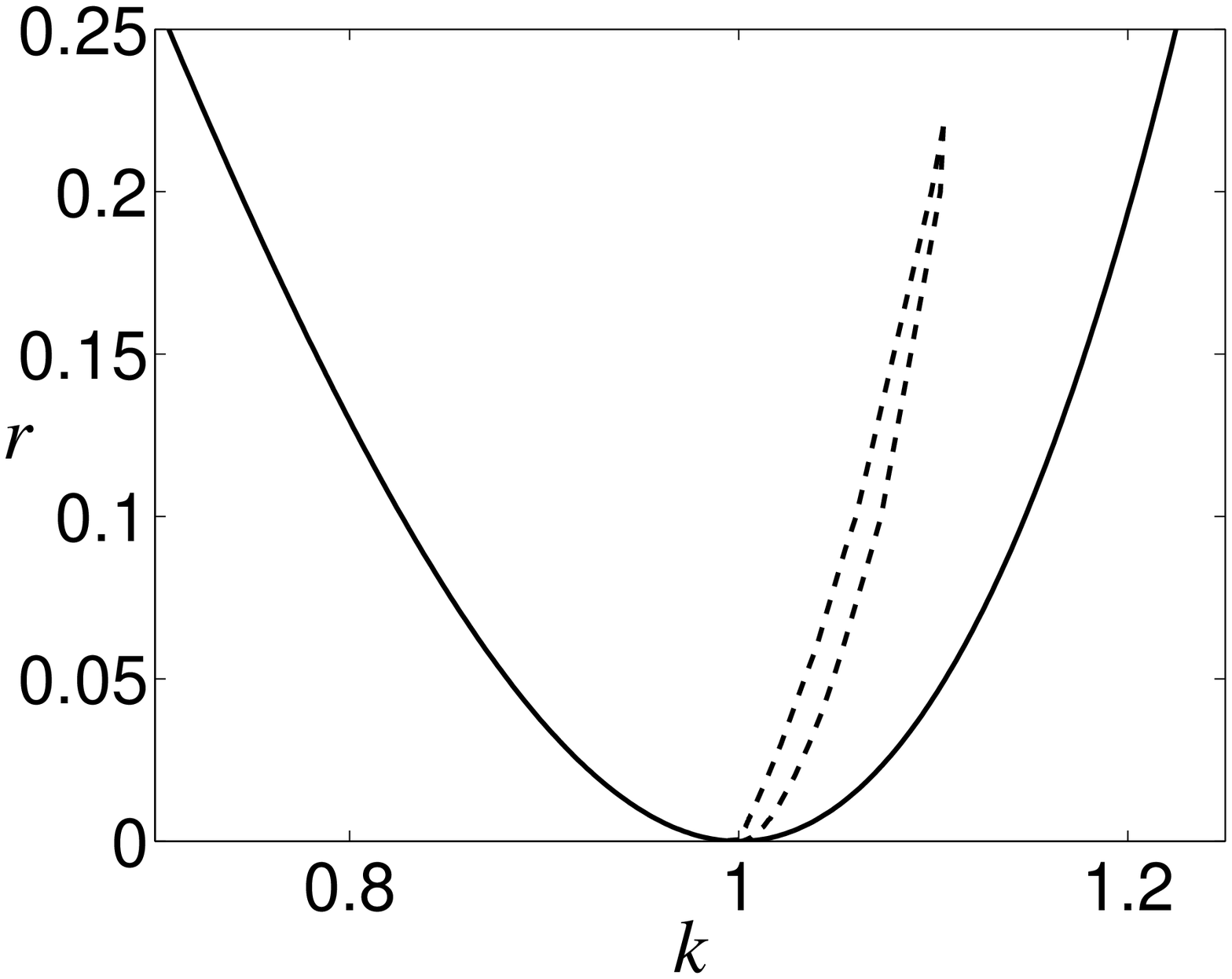}
(c)\includegraphics[height=0.2\linewidth]{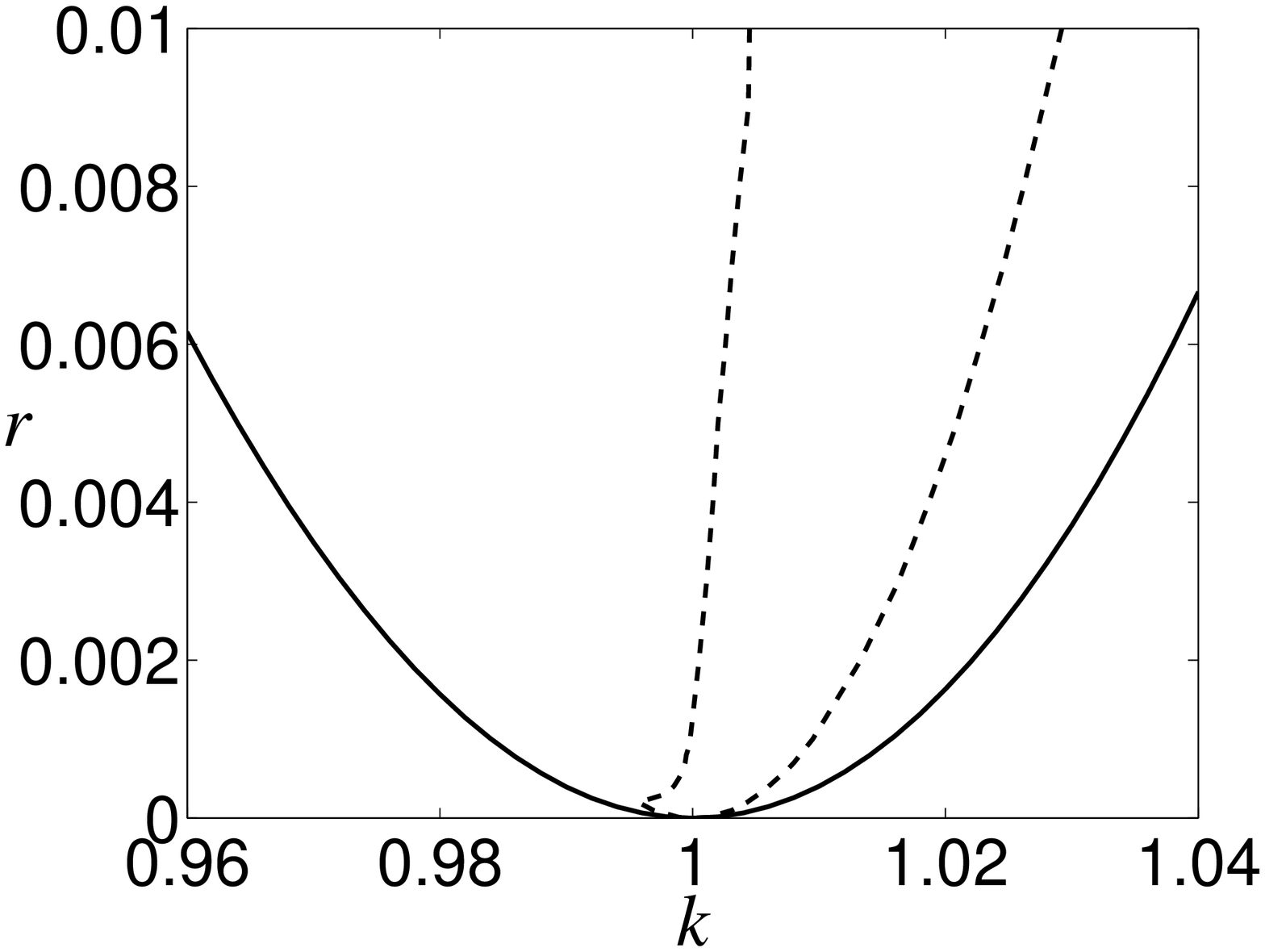}

\caption{The secondary stability regions of traveling waves of
\eqref{eq:nikd}, calculated numerically for (a)~$\alpha=1/2$, (b)~$\alpha=2$
and (c)~$\alpha=5$, all for $\beta=0$. Shown are the marginal
curve $r=(1-k^2)^2$ (solid line) and the secondary stability boundary of
the traveling waves (dashed line), with stability between the dashed
lines.}

\label{fig:32}
\end{figure}

\begin{figure}

(a)\includegraphics[height=0.2\linewidth]{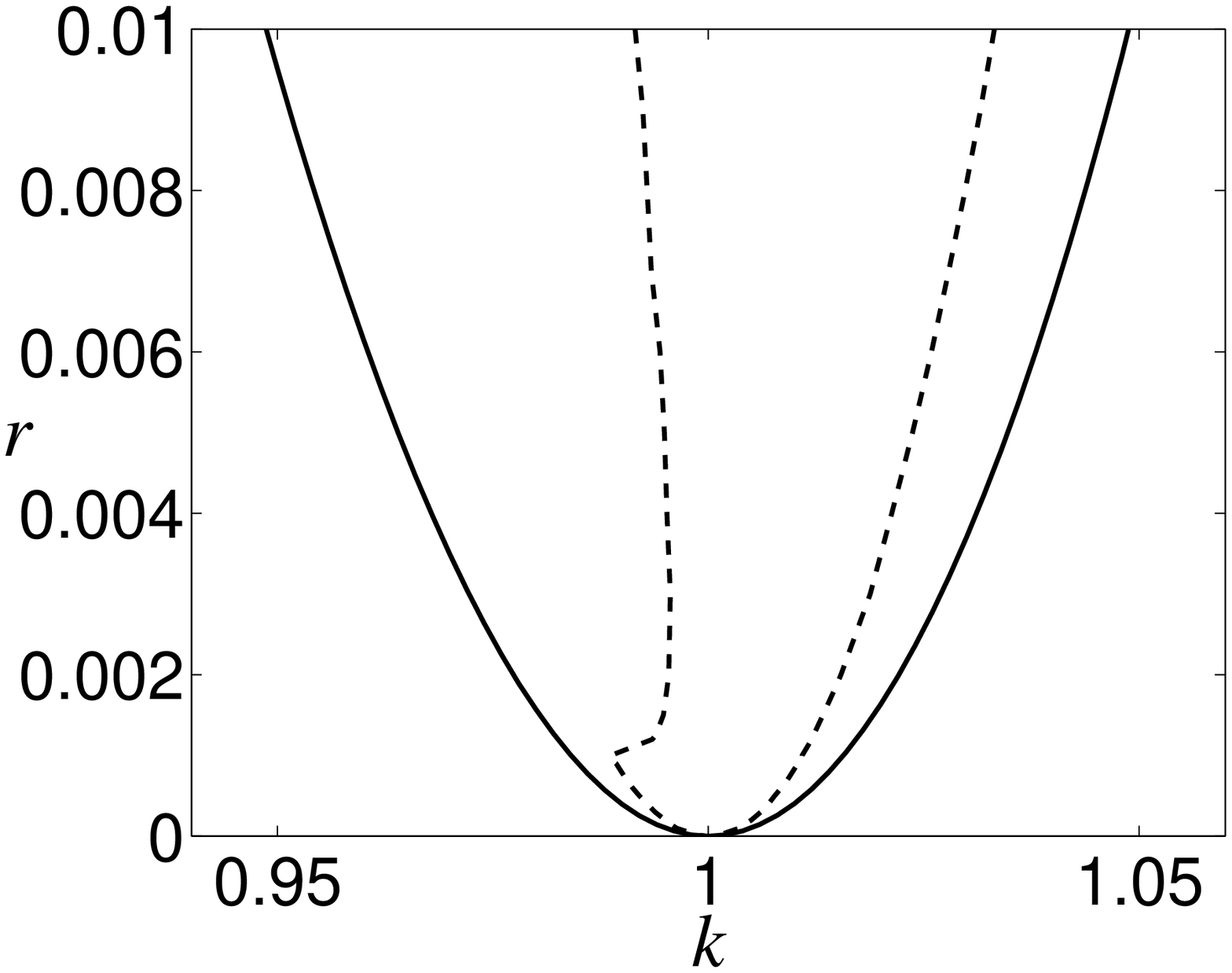}
(b)\includegraphics[height=0.2\linewidth]{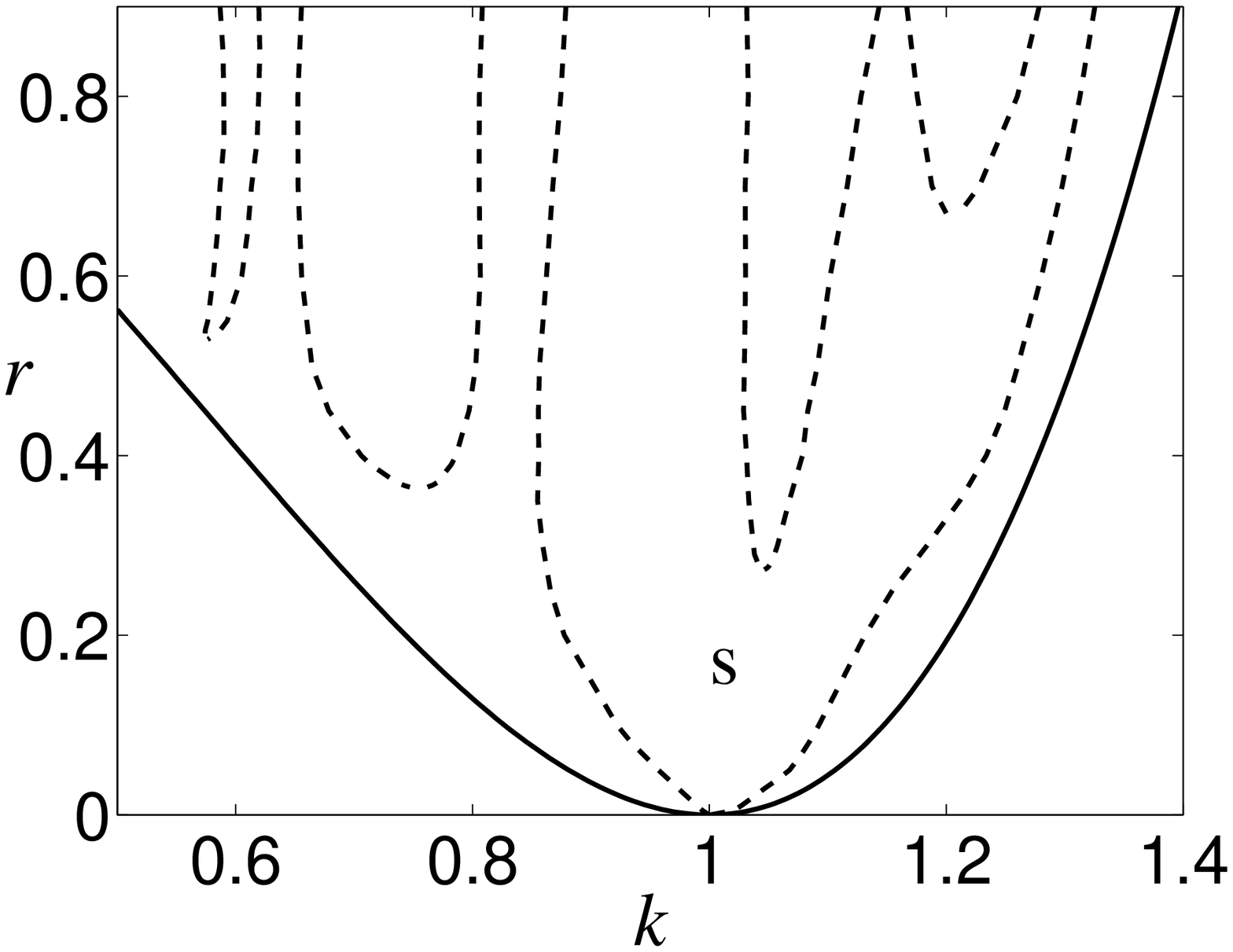}

\caption{The secondary stability regions of traveling waves of
\eqref{eq:nikd}, calculated numerically for (a)~$\beta=5$, (b)~$\beta=5.5$,
both with $\alpha=40$. Shown are the marginal curve
$r=(1-k^2)^2$ (solid line) and the secondary stability boundaries of the
traveling waves (dashed line), with stability between the dashed lines. To
clarify the regions of stability/instability, the ``s'' indicates one of
the stable regions.}

\label{fig:33}
\end{figure}

Now we present the secondary stability diagrams. The first case considered
here is setting $\beta=0$ and varying $\alpha$ --- see
Fig.~\ref{fig:32}. When $\alpha$ is small ($\alpha=1/2$), 
there is a very small region of stable waves in the $(k,r)$ plane.
The stable
region is a thin strip, confined to small values of $r$; in this case, for
$r>0.0078$ all rolls are unstable.

For larger $\alpha$ this strip of stable waves is longer and wider; for
example at $\alpha=2$ there are some stable rolls up to $r\approx0.22$,
and at $\alpha=5$ the stability region extends at least as far as $r=0.9$.
Furthermore, it is apparent for $\alpha=5$ that
a symmetrical Eckhaus-like stability region is
present for very small values of $r$ (from the numerical results
themselves, it seems to be present in all three
cases, but is visible only in the last plot of Fig.~\ref{fig:32}). The
shrinkage of the region of stable traveling waves for small $\alpha$ is
consistent with there being no stable rolls at all in the nondispersive
case.


While an exhaustive examination of the secondary stability diagrams across
$(\alpha,\beta)$ parameter space is infeasible, it is worthy of note that
these diagrams may be extremely complicated. A good example arises if we
set $\alpha=40$ and vary $\beta$  --- see Fig.~\ref{fig:33}. For $\beta=5$
there is a small Eckhaus-like stability region for $r<0.001$; for
larger values of $r$, there remains a single 
stability region. For larger $\beta$, however, the stability region splits
into several parts; for example, at $\beta=5.5$ there may be up to five
separate intervals of stable traveling waves for a given value of $r$.

\begin{figure}

\begin{center}
(a)\includegraphics[height=0.2\linewidth]{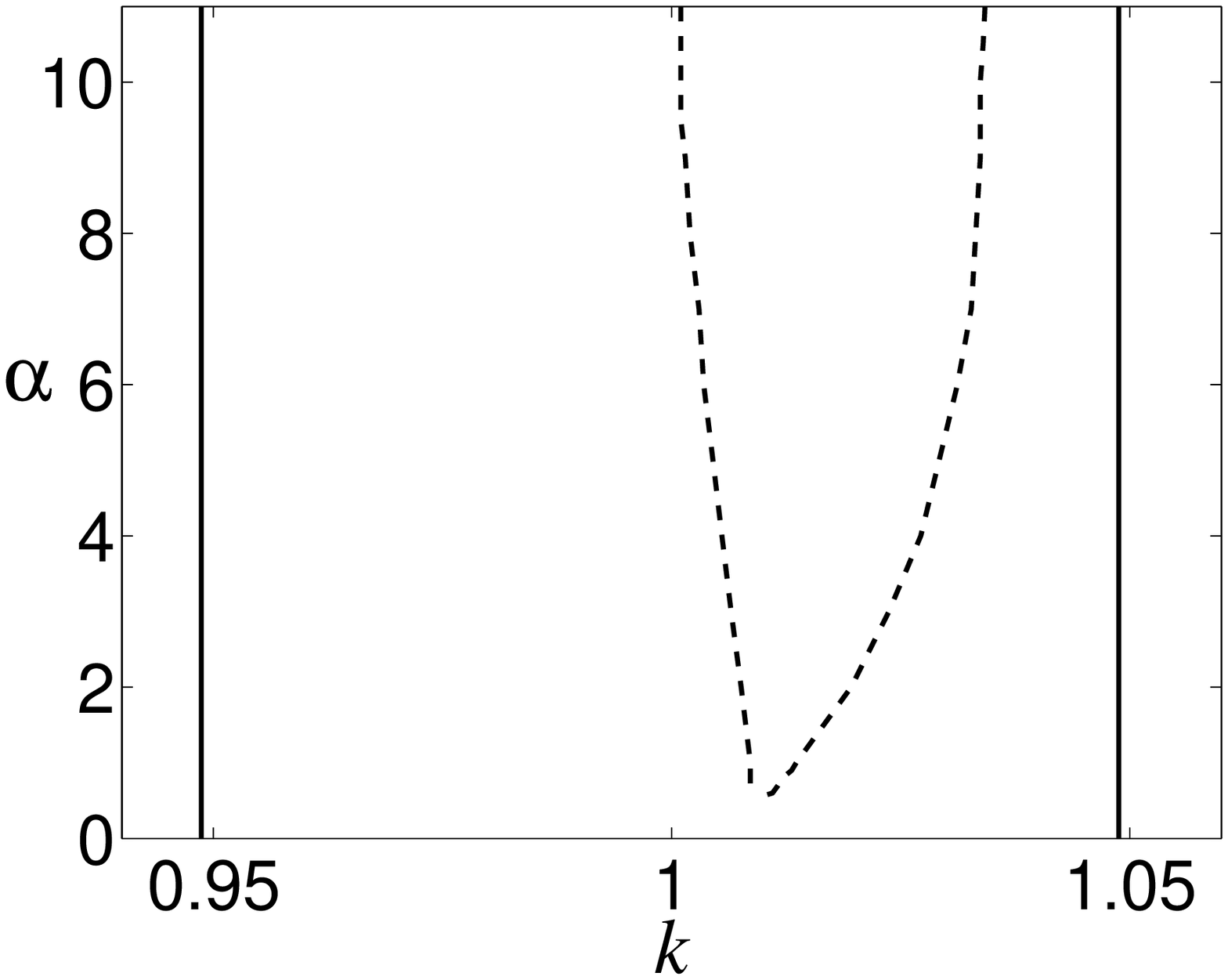}
(b)\includegraphics[height=0.2\linewidth]{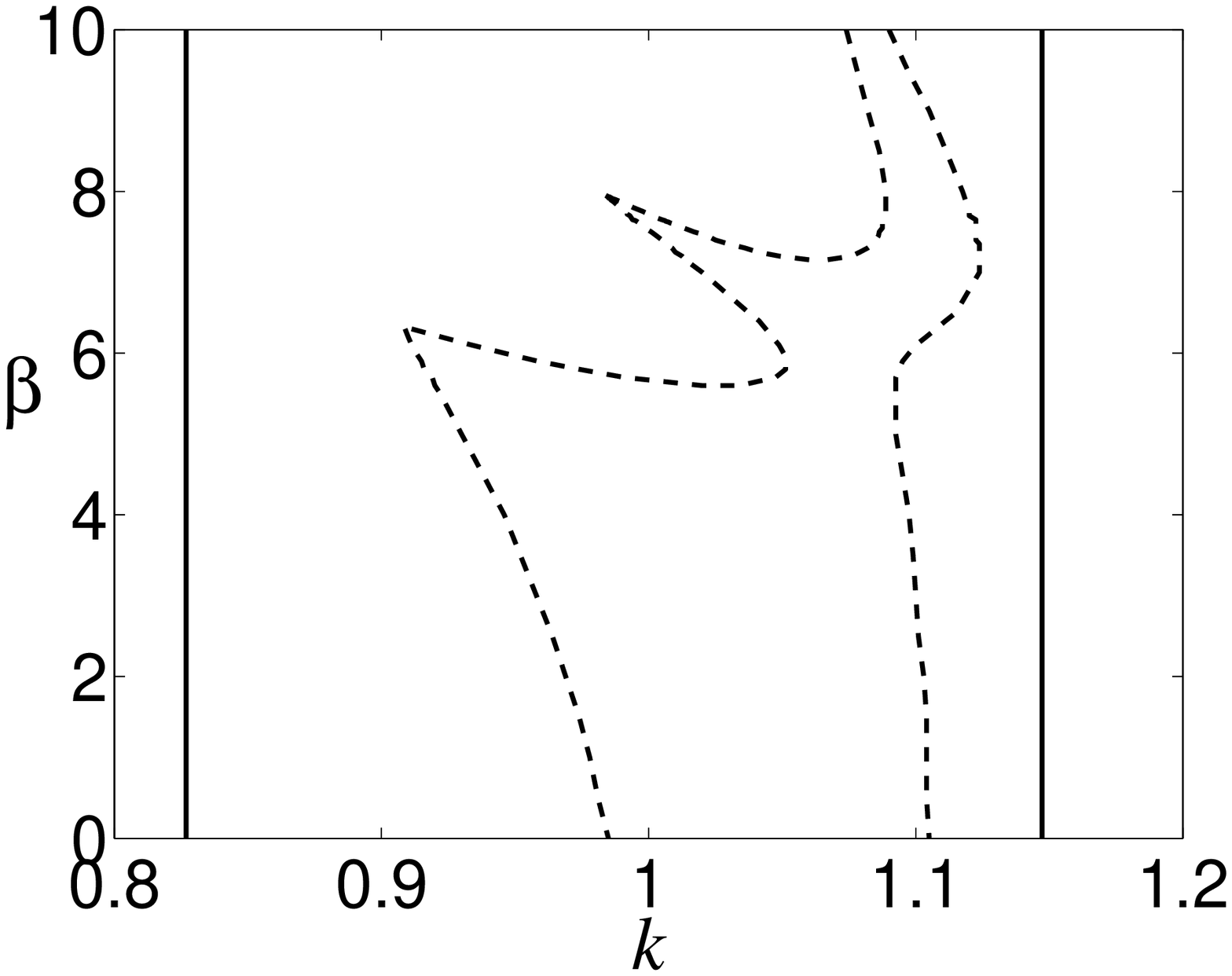}
\end{center}

\caption{The secondary stability of traveling waves of \eqref{eq:nikd}
calculated numerically for (a) fixed $\beta=0$ and $r=0.01$ in
$(k,\alpha)$ parameter space, (b) fixed $\alpha=40$ and $r=0.1$ in
$(k,\beta)$ parameter space. The marginal curve is represented by the
solid lines; traveling waves are stable inside the dashed lines.}

\label{fig:35}
\end{figure}

Above we have presented our secondary stability diagrams in the $(k,r)$
plane, for fixed values of $\alpha$ and $\beta$. If our interest is in the
effects of dispersion on the stability of traveling waves then it is more
instructive instead to fix $r$ and present results in either the $(k,
\alpha)$ or the $(k, \beta)$ plane. Our first example is for $r=0.01$ and
$\beta=0$ --- see Fig.~\ref{fig:35}(a). Given this value of $r$, the
traveling waves exist for $0.9487<k<1.0488$. We expect 
that if $\alpha$ is small enough then all roll
solutions are unstable; this is indeed the case. For larger values of
$\alpha$, a region of stable rolls appears. In Fig.~\ref{fig:35}(b), we
show a second case, where we fix $\alpha=40$ and $r=0.1$, to emphasize
that the structure of the stability region may be rather complicated,
exhibiting a sensitive parameter dependence.

\section{Secondary stability of traveling waves: $\alpha,\beta=O(1)$}
\label{sec:strong}

In this and the following two sections, we analyse the secondary stability of
traveling waves \eqref{eq:TW}.  The most straightforward case 
arises when the dispersion parameters $\alpha$ and $\beta$ 
are each $O(1)$. To contrast with later sections, we shall characterize 
this case as {\em strong dispersion}. Whereas the nondispersive 
Nikolaevskiy equation has no {\em stable} spatially periodic states, 
Kudryashov and Migita~\cite{KudMig} found stable periodic waves in their 
numerical simulations of the dispersive PDE \eqref{eq:nikd}, in this 
regime.

We begin by introducing the weakly nonlinear expansion
 \begin{equation}
u=\epsilon u_1+\epsilon^2 u_2+\epsilon^3 u_3+\cdots,
\label{eq:6}
 \end{equation}
with $r$ given by \eqref{eq:r2}. Then substitution in \eqref{eq:nikd} and
consideration of successive orders in $\ep$ leads to the following.

At $O(\ep)$, we find that
 \[
u_1=A\rme^{\rmi(x-c_0t)}+\cc,
 \]
where $c_0=\alpha-\beta$, and where the amplitude $A$ varies slowly in
space and in time, in principle depending on the slow variables
 \[
X=\ep x,\qquad \tau=\ep t,\qquad T=\ep^2 t.
 \]
A consideration of the terms proportional to $\rme^{\rmi(x-c_0t)}$ at
$O(\ep^2)$ then shows that in fact $A=A(\xi,T)$, where
 \[
\xi=X-(3\alpha-5\beta)\tau\equiv X-v\tau
 \]
is a coordinate moving at the group velocity of the waves. Then solving
the problem at this order in $\ep$ yields
 \[
u_2=-\frac{\rmi A^2}
{36(1+\rmi(\alpha-5\beta)/6)}\rme^{2\rmi(x-c_0t)}+\cc+f.
 \]
Here $f$ is a slow varying function of $X$, $\tau$ and $T$, chosen to 
appear at this order to balance forcing terms appearing at the next order 
in $\ep$.

At $O(\ep^3)$, we find, from the respective consideration of the terms in
\eqref{eq:nikd} proportional to $\rme^{\rmi(x-c_0t)}$ and
$\rme^{0\rmi(x-c_0t)}$, the amplitude equations
 \begin{eqnarray}
\pder{A}{T}&=&
\left(r_2-\frac{1-\rmi(\alpha-5\beta)/6}{36+(\alpha-5\beta)^2}|A|^2\right)A
+(4+\rmi(3\alpha-10\beta))\pder{^2A}{\xi^2}-\rmi fA,\label{eq:strAf}\\
\pder{f}{\tau}&=&-\pder{|A|^2}{\xi}.\label{eq:strf}
 \end{eqnarray}

Since $A=A(\xi,T)$, the second amplitude equation suggests
taking $f=f(\xi,T)$, in which case \eqref{eq:strf} becomes
 \[
-v \pder{f}{\xi}=-\pder{|A|^2}{\xi},
 \]
and hence $vf=|A|^2+K(T)$, for some $K(T)$. However, the constraint that
the spatial average of $u$ should be zero gives
$K(T)=-\langle|A|^2\rangle$, where the angle brackets denote the average
in $\xi$. Thus
 \beq
f=\frac{-\langle|A|^2\rangle+|A|^2}{v}
 \label{eq:strfis}
 \eeq
and the amplitude equation \eqref{eq:strAf} becomes the nonlocal
Ginzburg--Landau equation
 \begin{equation}
\pder{A}{T}=
\left(r_2-\frac{1-\rmi(\alpha-5\beta)/6}{36+(\alpha-5\beta)^2}|A|^2
+\rmi\frac{\langle|A|^2\rangle-|A|^2}{v}\right)A+
(4+\rmi(3\alpha-10\beta))\pder{^2A}{\xi^2}.
\label{eq:strA}
 \end{equation}
It is worth mentioning that in view of \eqref{eq:strfis} the present
scaling breaks down when $v$ is small; in particular, this is the case
when $\alpha$ and $\beta$ are both small, and this case will be
considered in later sections.

It is helpful in analysing \eqref{eq:strA} to put it in canonical form by 
rescaling all the variables, to give
 \begin{equation}
\pder{A}{T}=A+\rmi d(\langle|A|^2\rangle-|A|^2)A+
(1+\rmi a)\pder{^2A}{\xi^2}-(1+\rmi b)|A|^2A, \label{eq:strAcanon}
 \end{equation}
where
 \[
a=\frac{3\alpha-10\beta}{4},\qquad
b=\frac{5\beta-\alpha}{6},\qquad
d=\frac{36+(5\beta-\alpha)^2}{v}.
 \]
Equations similar to \eqref{eq:strAcanon}, including a nonlocal nonlinear 
term have been derived and studied in the context of convection in a 
rotating annulus~\cite{PlautBusse} and in electrical and magnetic 
systems~\cite{DLT,Elmer}.

\begin{figure}
\includegraphics[width=0.6\linewidth]{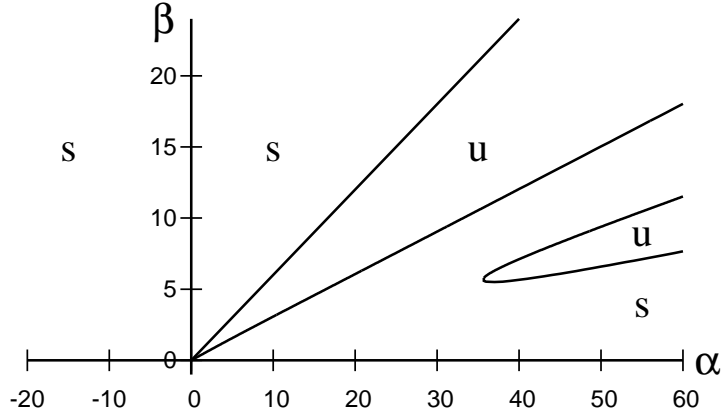}

\caption{Diagram showing the sign of $1+a(b+d)$ in $\alpha\beta$ parameter
space. Regions with ``s'' are where $1+a(b+d)>0$, so that a limited band
of plane waves is stable, as in \eqref{eq:qc}; a ``u'' indicates where
$1+a(b+d)<0$, and all plane waves are unstable.}

\label{fig:+-}
\end{figure}

Equipped with \eqref{eq:strAcanon}, we are now in a position to explore
the secondary stability of weakly nonlinear spatially periodic solutions
of the dispersive Nikolaevskiy equation. Such solutions correspond to
plane-wave solutions of \eqref{eq:strAcanon}, which exist in the form $A=P
e^{\rmi(\omega T+q\xi)}$, with $P=(1-q^2)^{1/2}$ and $\omega=q^2(b-a)-b$.
To study the stability of the plane-wave solution, we write
$A=(1+p(\xi,T))P e^{\rmi(\omega T+q\xi)}$, which, after substitution in
\eqref{eq:strAcanon} and linearization in the perturbation $p$, yields
 \[
\pder{p}{T}=(1+\rmi a)\left(\pder{^2p}{\xi^2}+2\rmi q\pder{p}{\xi}\right)
-(1+\rmi b)P^2(p^*+p)+\rmi dP^2\left(\langle p+p^*\rangle-(p+p^*)\right).
 \]
Then upon setting $p(\xi,T)=R(T)\rme^{\rmi L\xi}+S^*(T)\rme^{-\rmi L\xi}$
and equating the coefficients of $\rme^{\rmi L\xi}$ and $\rme^{-\rmi
L\xi}$, we have
 \bean
\deriv{R}{T}&=&-(1+\rmi a)L(LR+2qR)-(1+\rmi b)Q^2(R+S)-\rmi dQ^2(R+S),\\
\deriv{S}{T}&=&-(1-\rmi a)L(LS-2qS)-(1-\rmi b)Q^2(R+S)+\rmi dQ^2(R+S).
 \eean
Finally, with $R(T)$ and $S(T)$ proportional to $\rme^{\mu T}$, and
expanding the growth rate in powers of the perturbation wavenumber $L$, we
have the dispersion relation
 \beq
\mu=-2\rmi q(a-b-d)L+L^2P^{-2}
\left(-1-a(b+d)+q^2\left[3+2(b+d)^2+a(b+d)\right]\right)+O(L^3).
 \label{eq:mu}
 \eeq

If we suppose (as is generally the case) that $a\neq b+d$ then it is 
apparent from \eqref{eq:mu} that the solution has a long-wavelength 
oscillatory instability whenever
 \[
q^2\left(3+2(b+d)^2+a(b+d)\right)>1+a(b+d).
 \]
Since $1+a(b+d)<3+a(b+d)+2(b+d)^2$, we see that stability is determined by
the following. If $1+a(b+d)>0$ then $3+a(b+d)+2(b+d)^2>0$, and the
plane-wave solutions are stable provided
 \beq
0\leq q^2<q_c^2\equiv\frac{1+a(b+d)}{3+2(b+d)^2+a(b+d)}<1.
 \label{eq:qc}
 \eeq
If instead $1+a(b+d)<0$, then all plane waves are unstable. We note that
setting $a=b=d=0$ reduces \eqref{eq:strAcanon} to a real Ginzburg--Landau
equation for $A$, and our results reduce to the usual Eckhaus instability
(with stability for $q^2<1/3$)~\cite{Eck,Hoy}.

To apply this result to (\ref{eq:nikd}) it is necessary to indicate the 
regions in $\alpha$, $\beta$ parameter space in which the quantity 
$1+a(b+d)$ is positive or negative. In Fig.~\ref{fig:+-}, regions 
denoted by ``s'' indicate where $1+a(b+d)>0$, so that there are some 
stable plane waves, as in \eqref{eq:qc}; those regions denoted by ``u'' 
show where $1+a(b+d)<0$, and hence all plane waves are unstable. As 
discussed in Sec.~\ref{sec:nik}, only the region $\beta \geq 0$ need be 
presented.
The existence of a stable region in Fig.~\ref{fig:+-} is consistent
with the numerical results of Sec.~\ref{sec:secstab}; for example
Fig.~\ref{fig:32} shows a stable region when $\alpha=O(1)$ and
$\beta=0$.

\begin{figure}
(a)
\includegraphics[width=0.25\linewidth]{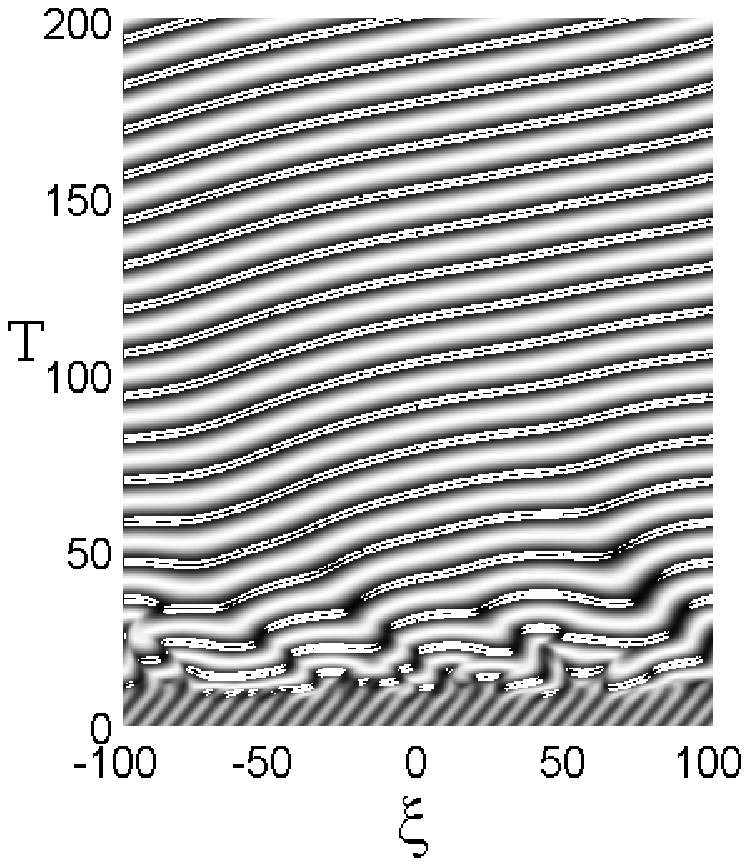}
(b)
\includegraphics[width=0.25\linewidth]{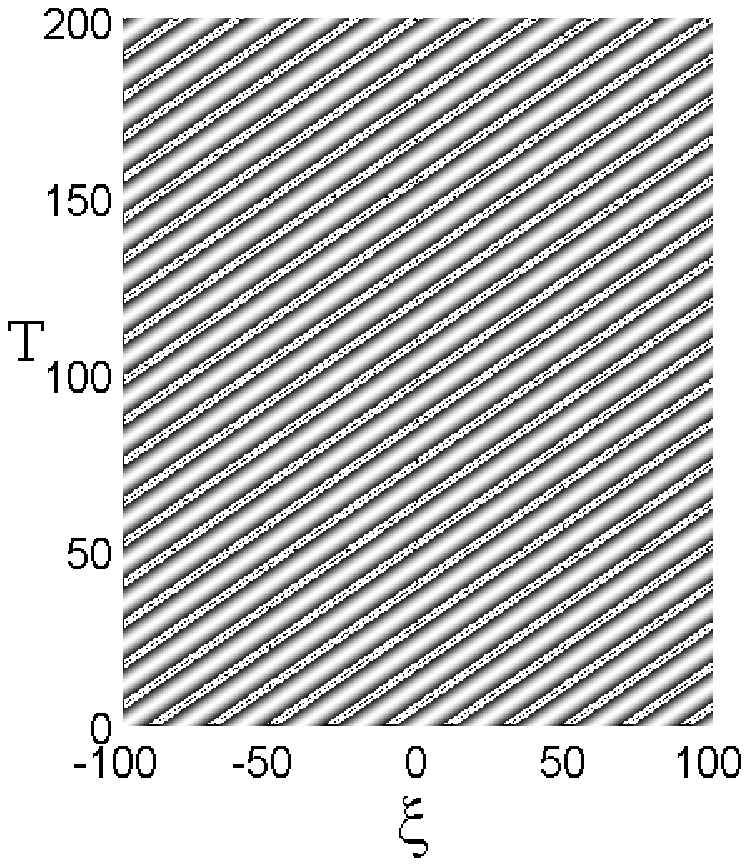}
(c)
\includegraphics[width=0.25\linewidth]{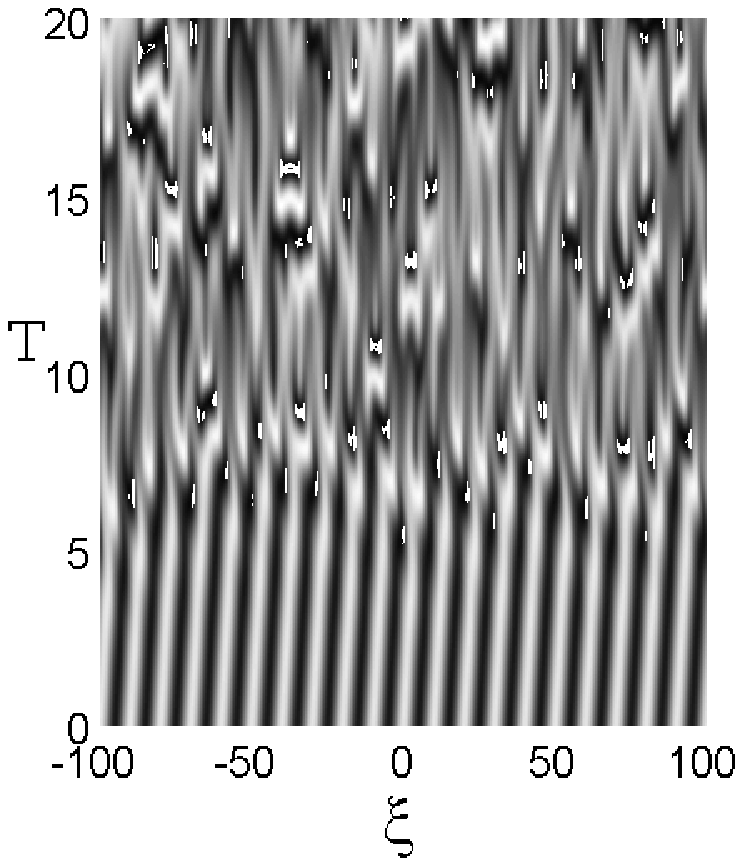}

\caption{Numerical simulations of the amplitude equation 
(\ref{eq:strAcanon}): in each case the real part of $A$ is plotted as a 
function of $\xi$ and $T$. In (a) and (b) $\alpha=10$, whereas in (c) 
$\alpha=8.4$; in each case $\beta=2.6$. The initial condition in each case 
is a plane wave, with $n$ wavelengths in the computational box 
$-32\pi<\xi<32\pi$, plus small-amplitude random noise: (a) $n=28$ (hence 
$q=0.875$); (b) $n=10$ ($q=0.3125$); (c) $n=20$ ($q=0.625$).}

\label{fig:strnum}
\end{figure}

In Fig.~\ref{fig:strnum}, we illustrate the considerations above with 
some numerical simulations of the modified complex Ginzburg--Landau 
equation \eqref{eq:strAcanon}. Our numerical code is pseudospectral, and 
uses exponential time differencing~\cite{CoxMatt}. In each case the 
initial condition is a plane wave plus small-amplitude random noise. For 
the simulations illustrated in Fig.~\ref{fig:strnum}(a) and~(b), 
$1+a(b+d)>0$. The two plots show the fate of initial conditions in the 
unstable and stable regions of Fig.~\ref{fig:+-}, respectively. In each 
case, a stable plane wave is obtained at large $T$. 
Figure~\ref{fig:strnum}(c) shows the development of instability in the 
case $1+a(b+d)<0$, where all plane waves are unstable. Here
the solution is persistently time-dependent.

The analysis above tells us about the secondary stability of 
traveling-wave solutions of the dispersive Nikolaevskiy equation when 
$\alpha,\beta=O(1)$, and the results are summarized in 
Fig.~\ref{fig:+-}. We may think of this analysis as holding for any 
fixed $\alpha$ and $\beta$ (not both zero) in the limit as $r\to0$; thus 
we expect the lowest part of the secondary stability diagram in $(r,k)$ 
parameter space to reflect Fig.~\ref{fig:+-}.

However, as indicated earlier, when $\alpha$ and $\beta$ are both small, 
the analysis above does not hold, and requires reconsideration. We should 
expect such analysis to break down in this limit, because 
Fig.~\ref{fig:+-} is inconsistent with the known behavior of the 
nondispersive Nikolaevskiy equation ($\alpha=\beta=0$), for which all 
rolls are unstable at onset~\cite{TribVel,MattCox00}. Thus in the next 
section we consider smaller values of $\alpha,\beta$.

\section{Secondary stability of traveling waves:
$\alpha,\beta=O(\epsilon^{3/4})$}
\label{sec:int}

It turns out, after some experimentation, that small $\alpha$ and $\beta$ 
first lead to a new scaling if we adapt the scaling first used by 
Tribelsky and Velarde~\cite{TribVel} for the nondispersive case, and 
extended by Cox and Matthews~\cite{CMdamp} to a damped version of the 
Nikolaevskiy equation. In this scaling the original traveling waves 
remain $O(\ep)$, but the perturbation to the traveling-wave amplitude is 
$O(\ep^{3/2})$ and the large-scale mode is $O(\ep^{7/4})$; furthermore, 
slow spatial and temporal variations of perturbations take place on scales 
given by $X=\epsilon^{3/4}x$, $T=\epsilon^{3/2}t$, $\tau=\epsilon^{3/4}t$. 
(Note that these slow variables are different from those of the previous 
section, but our notation for slow variables is consistent within 
sections.) To allow the development of consistent amplitude equations for 
the perturbation we then take
 \[
\alpha=\ep^{3/4}\hat{\alpha},\qquad
\beta=\ep^{3/4}\hat{\beta}.
 \]

Applying a weakly nonlinear analysis to \eqref{eq:nikd} gives
 \beq
u=\ep(a_0+\ep^{1/2}a(X,T))\rme^{\rmi M}+\cc+\ep^{7/4}f(X,T)+\cdots,
 \label{eq:u}
 \eeq
where $a_0=6\sqrt{r_2-4q^2}$,
 \[
M=(1+\epsilon q)x-\hat c\tau-\ep^{1/4}\hat vqT+\ep^{1/4}\psi(X,T),
 \] 
$\hat c=\hat{\alpha}-\hat{\beta}$ and $\hat v=3\hat{\alpha}-5\hat{\beta}$. 
Here $a(X,T)$ represents disturbances to the amplitude of the pattern, 
$\psi(X,T)$ represents corresponding disturbances to the phase of the 
pattern and $f(X,T)$ is a large-scale mode. Substitution of $u$, as given 
by \eqref{eq:u}, in \eqref{eq:nikd} requires the consideration of the 
problem at successive orders in $\ep^{1/4}$. After much consequent 
algebra, we find the (nonlinear) amplitude equations
 \begin{eqnarray*}
\pder{\psi}{T}&=&4\pder{^2\psi}{X^2}-f-\hat v\pder{\psi}{X},\\
\pder{f}{T}&=&\pder{^2 f}{X^2}-2a_0\pder{a}{X},\\
\pder{a}{T}&=&4\pder{^2 a}{X^2}-
4a_0\left(\pder{\psi}{X}\right)^2-8a_0q\pder{\psi}{X}-\hat v\pder{a}{X}.
 \end{eqnarray*}
Note that dispersion is represented in these equations only through the 
terms $\hat v\psi_X$ and $\hat va_X$, representing advection of the 
pattern envelope with the group velocity $\hat v$. Note also that the group 
velocity of the large scale mode $f$ is zero, and hence no corresponding 
term appears in the second of these equations.

The three amplitude equations may be reduced to the single (nonlinear)
phase equation
 \begin{equation}
\left(\pder{}{T}-4\pder{^2}{X^2}+\hat v\pder{}{X}\right)^2
\left(\pder{}{T}-\pder{^2}{X^2}\right)\psi=
-16a_0^2\left(\pder{\psi}{X}+q\right)\pder{^2{\psi}}{X^2}.
\label{eq:11}
 \end{equation}
Then linearising this equation and setting $\psi=e^{\rmi LX+\sigma T}$
yields the dispersion relation
 \begin{equation}
\sigma^3+9\sigma^2L^2
+24\sigma L^4
-\hat v^2\sigma L^2
+16L^6
-\hat v^2L^4
-16a_0^2qL^2
+\rmi \hat v(2\sigma^2L+10\sigma L^3+8L^5)=0.
\label{eq:d}
 \end{equation}
Before considering this dispersion relation for general $L$, it is helpful 
to consider the two limiting cases, of small and large $L$. First, if $L$ 
is small, then $\sigma^3\sim16a_0^2qL^2$. Thus, to leading order in $L$, 
$\sigma=\sigma_{2/3}L^{2/3}$, where $\sigma^3_{2/3}=16a_0^2q$; hence all 
traveling waves are unstable if $L$ is small. On the other hand, if $L$ 
is large, then we have $\sigma^3+9\sigma^2L^2+24\sigma L^4+16L^6\approx0$, 
and so $\sigma\approx-L^2$ or $-4L^2$ (twice); hence traveling waves are 
stable to large-$L$ disturbances. In summary, all traveling waves are 
unstable at onset (provided $a_0^2q\neq0$; in fact we shall see later that 
when $a_0^2q$ is suitably small, we shall need to reconsider this 
conclusion). The rest of the section provides more details of the 
instability, for general values of $L$.

In order to find the secondary stability boundary for the traveling 
waves, we set $\sigma=\rmi\Omega$ in the dispersion relation \eqref{eq:d}, 
where $\Omega$ is real. From the real and the imaginary parts, we obtain
 \begin{eqnarray*}
\Omega^2-\frac{16}{9}L^4+\frac{16}{9}a_0^2q+\frac{\hat v^2}{9} L^2
+\frac{10}{9}\hat vL\Omega&=&0,\\
\Omega^3-24\Omega L^4+\hat v^2\Omega L^2+2\hat vL\Omega^2-8\hat vL^5&=&0,
 \end{eqnarray*}
and then after eliminating $\Omega$ between these two equations we find
that this stability boundary is given by
 \begin{equation}
16a_0^6q^3-2500L^{12}+2100L^8a_0^2q+384L^4a_0^4q^2-
200\hat v^2L^{10}-4\hat v^4L^8-
44\hat v^2L^6a_0^2q+\hat v^2L^2a_0^4q^2=0.
\label{eq:4}
 \end{equation}
We note that in this equation $L$ and $\hat v$ appear only as even powers 
and thus we can restrict our attention to positive $L$ and $\hat v$ with 
no loss of generality. However, both even and odd powers of $q$ occur, so 
no such economy is possible in considering $q$ (indeed, in the light 
of~\cite{TribVel}, we should expect different behaviors for $q>0$ and 
$q<0$).

\begin{figure}
(a)
\includegraphics[width=0.4\linewidth]{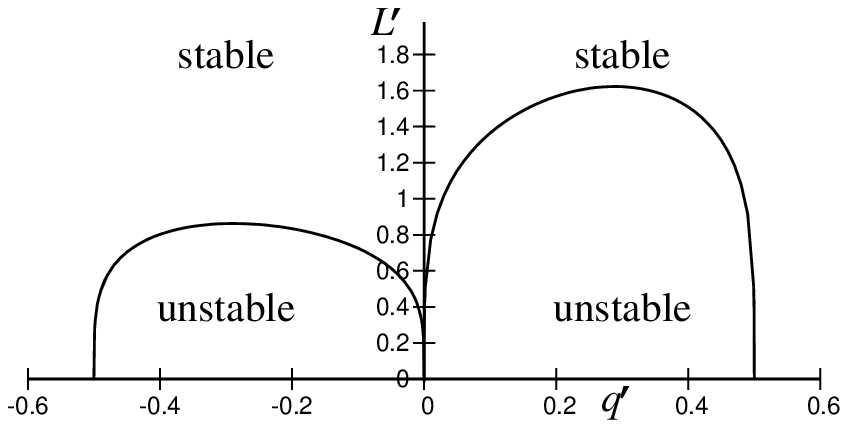}
(b)
\includegraphics[width=0.4\linewidth]{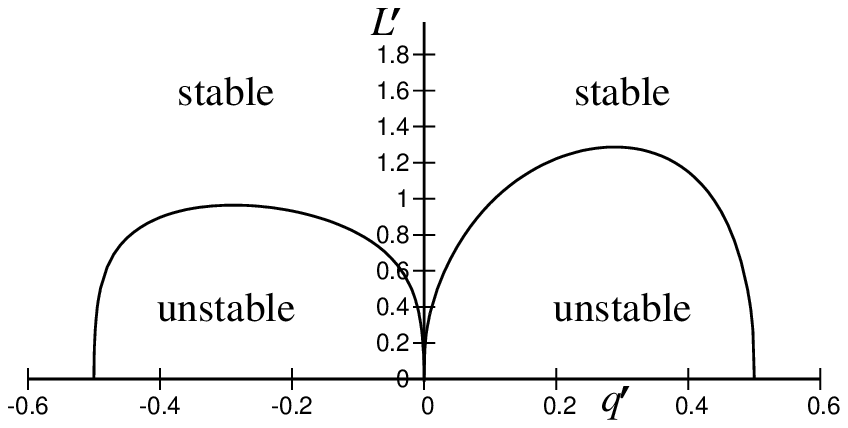}

\caption{Predicted secondary stability boundaries of spatially periodic
solutions of the Nikolaevskiy equation. (a)
Nondispersive case ($v'=0$), (b) dispersive case ($v'=5$). }
\label{fig:1}
\end{figure}

For the case of no dispersion, Tribelsky and Velarde~\cite{TribVel} showed 
that, according to the present scaling, there is monotonic instability of 
the rolls with $q>0$ (with unstable disturbances having 
$0<L<(a_0^2q)^{1/4}$). By contrast, oscillatory instability occurs for 
rolls with $q<0$ (unstable modes having $0<L<(-2a_0^2q/25)^{1/4}$).

It is convenient to present our results for the dispersive case in terms 
of the rescaled variables $q'=q/r_2^{1/2}$, $L'=L/r_2^{3/8}$ and $v'=\hat 
v/r_2^{3/8}$. Figure~\ref{fig:1} illustrates the regions of stability and 
instability of the traveling waves, in the cases $v'=0$ and $v'=5$. Note 
that in the dispersive case all instabilities are oscillatory.

We should view with caution the conclusion above that all traveling waves
are unstable, because it relies crucially on the assumption that $a_0^2q$
is not small. The stability analysis above breaks down if $q$ or $a_0^2$
are small; the true stability properties of corresponding traveling waves
will be investigated in the next section.

\section{Secondary stability of traveling waves: $\alpha,\beta=O(\ep)$}

In this section, we investigate the cases of traveling waves with
wavenumber close to $k=1$ or close to the marginal stability boundary, in
other words those for which in the previous scaling $a_0^2q\ll1$.

\subsection{Traveling waves with close-to-critical wavenumber}
\label{sec:3}

In order to resolve the secondary stability problem for traveling waves
with wavenumber close to $k_c=1$, we set $k=1+\epsilon^2 q$, as was done
for the nondispersive case by Tribelsky and Velarde~\cite{TribVel}. Then a
distinguished balance occurs for $\alpha,\beta=O(\ep)$; so we write
 \[
\alpha=\ep\hat{\alpha},\qquad
\beta=\ep\hat{\beta}.
 \]
Upon setting $\hat c=\hat{\alpha}-\hat{\beta}$, $r=\ep^2 r_2$, $\hat
v=3\hat{\alpha}-5\hat{\beta}$, $X=\ep x$, $\tau=\ep t$ and $T=\ep^2 t$
(the scalings for $X$ and $T$ being as in~\cite{TribVel}), we find from
\eqref{eq:nikd} that
 \[
u=\ep(6\sqrt{r_2}+\ep^2 a(X,T))\rme^{\rmi M}+\cc+\ep^3f(X,T)+\cdots,
\]
where now
 \[
M=(1+\ep^2 q)x-c\tau+
\ep(-\hat vq+\tsfrac16 r_2(\hat{\alpha}-5\hat{\beta}))T+\ep\psi(X,T).
 \]
The terms in $M$ involving $\hat{\alpha}$ and $\hat{\beta}$ correspond to
nonlinear effects of the finite traveling-wave amplitude on the speed
of the waves; see~\eqref{eq:TWspeed}.

After much algebra, the relevant (nonlinear) amplitude equations are found
to be, at $O(\ep^4)$ and $O(\ep^5)$,
 \bea
\pder{\psi}{T}&=&4\pder{^2 \psi}{X^2}-f-\hat v\pder{\psi}{X},\label{eq:P}\\
\pder{f}{T}&=&\pder{^2 f}{X^2}-12r_2^{1/2}\pder{a}{X},\\
\pder{a}{T}&=&4\pder{^2a}{X^2}-24r_2^{1/2}\left(\pder{\psi}{X}\right)^2
-\hat v\pder{a}{X}-6r_2^{1/2}\pder{f}{X}-2r_2a\nonumber\\
&&{}+6r_2^{1/2}\left(-8q+\frac{22}{3}r_2+12\pder{^2}{X^2}+
(10\hat{\beta}-3\hat{\alpha})\pder{}{X}\right)\pder{\psi}{X}.\label{eq:a}
 \eea
We note that in these equations the influence of dispersion arises not 
only through the terms involving the group velocity $\hat v$, but also 
through the term $10\hat{\beta}-3\hat{\alpha}$ in the equation for $a_T$, 
in contrast to the previous case.

To determine the stability of the traveling waves, these equations are
linearized; for solutions proportional to $\rme^{\rmi LX+\sigma T}$, we
find the dispersion relation
 \bea
&&\sigma^3+9\sigma^2L^2+24\sigma L^4+16L^6+528r_2^2L^2\nonumber\\
&&{}-576r_2qL^2+82r_2\sigma L^2-568r_2L^4+2r_2\sigma^2-\hat v^2\sigma 
L^2-L^4\hat v^2 \nonumber \\
&&{}+\rmi(2r_2\hat v\sigma L+360r_2\hat{\beta}
L^3+8\hat vL^5+10\hat v\sigma L^3+2\hat v\sigma^2L+2r_2\hat vL^3)=0.
\label{eq:8}
 \eea
As in the previous section, in the limit of large $L$, all eigenvalues
have negative real part. By contrast, in the limit of small $L$, if we
expand $\sigma=\sigma_1L+\sigma_2L^2+\cdots$, then from \eqref{eq:8} we
find that $\sigma_1$ satisfies
 \[
r_2\sigma_1^2-288r_2q+264r_2^2+ir_2\hat v\sigma_1=0,
 \]
whereas $\sigma_2$ is determined from
 \[
\sigma_1^3+82r_2\sigma_1+4r_2\sigma_1\sigma_2-\hat v^2\sigma_1
+2\rmi(r_2\hat v\sigma_2+\hat v\sigma_1^2+r_2\hat v+180r_2\hat{\beta})=0.
 \]
The first of these gives
 \begin{equation}
\sigma_1=\frac{1}{2}\left(-\rmi \hat v\pm\sqrt{-\hat v^2+1152q-1056r_2}\right),
\label{eq:sig1}
 \end{equation}
and so traveling waves are certainly unstable if their wavenumber 
satisfies $q>11r_2/12+\hat v^2/1152$. The term $\hat v^2/1152$ indicates 
that these waves become more stable with respect to this instability in 
the presence of dispersion. If instead $q<11r_2/12+\hat v^2/1152$, then 
$\sigma_1$ is purely imaginary, and stability is determined by
 \beq
\sigma_2=
\pm\frac{-72 \hat vq/r_2+171 \hat v/2
-180\hat{\beta}}{\sqrt{\hat v^2-1152q+1056r_2}}+
\frac{91}{2}-\frac{72}{r_2}q,
 \label{eq:sig2}
 \eeq
a consideration of which shows that these waves are made more unstable
to the long-wavelength oscillatory instability in the presence of dispersion.

\begin{figure}
(a)\includegraphics[width=0.25\linewidth]{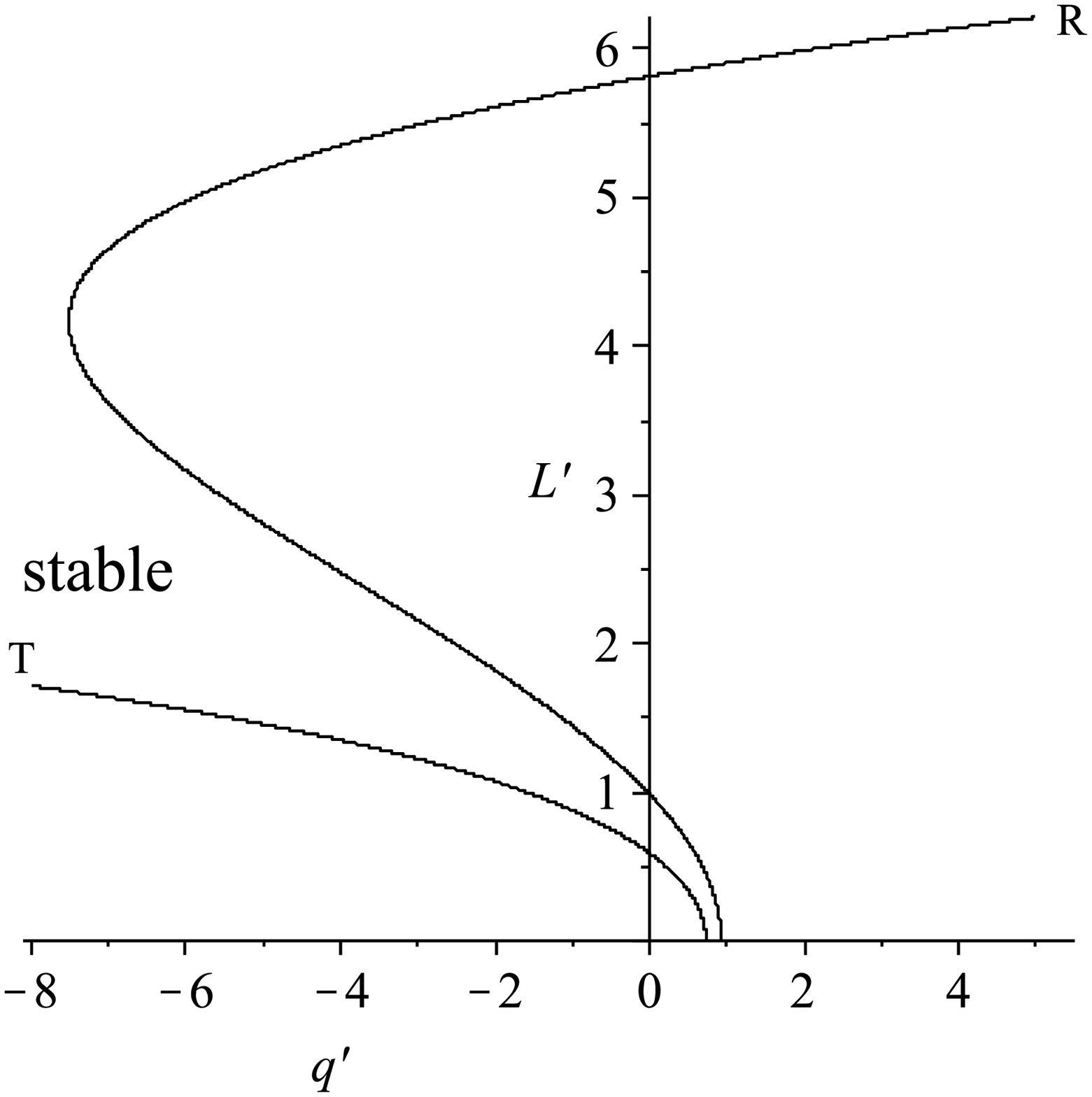}
(b)\includegraphics[width=0.25\linewidth]{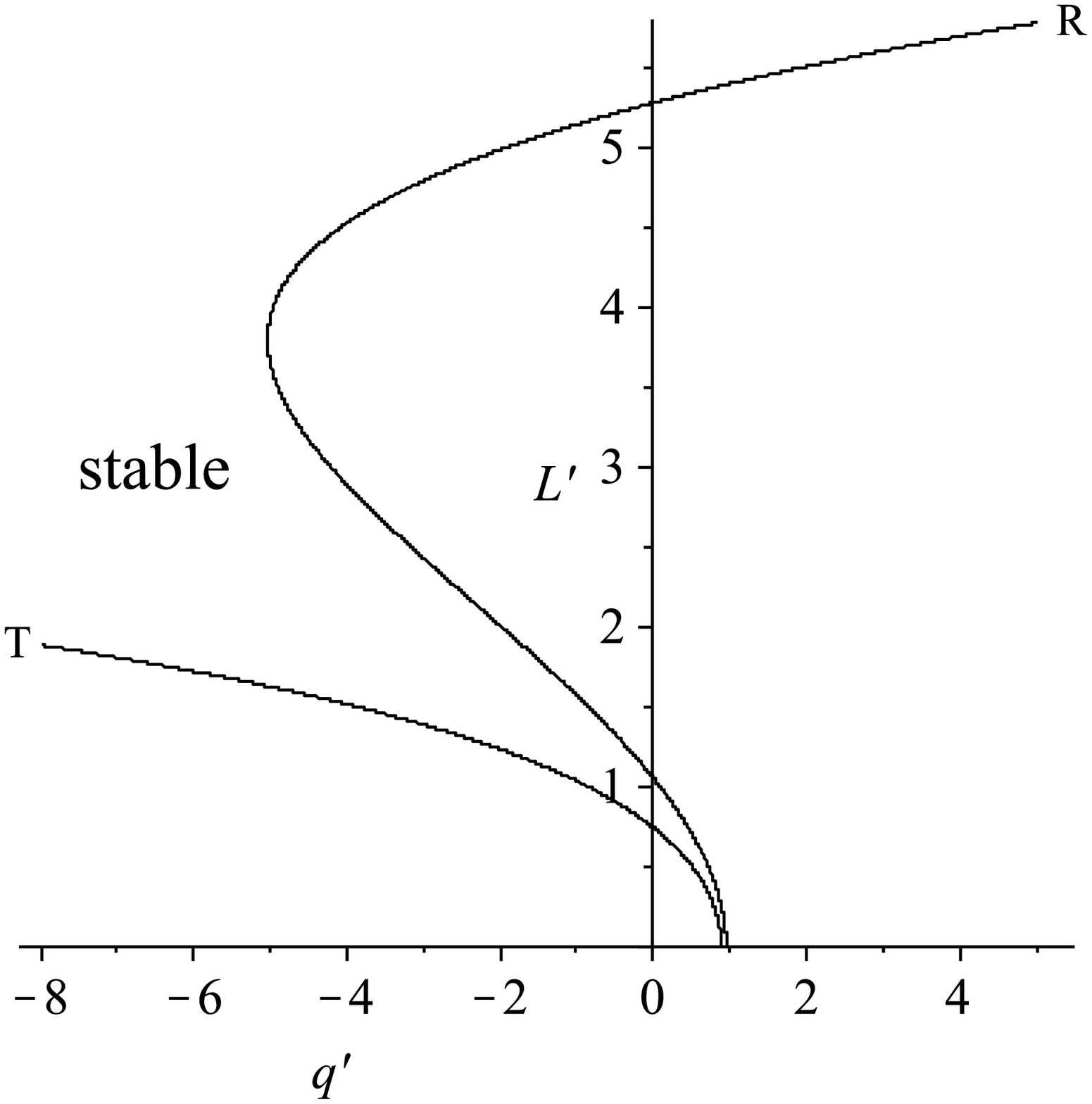}
(c)\includegraphics[width=0.25\linewidth]{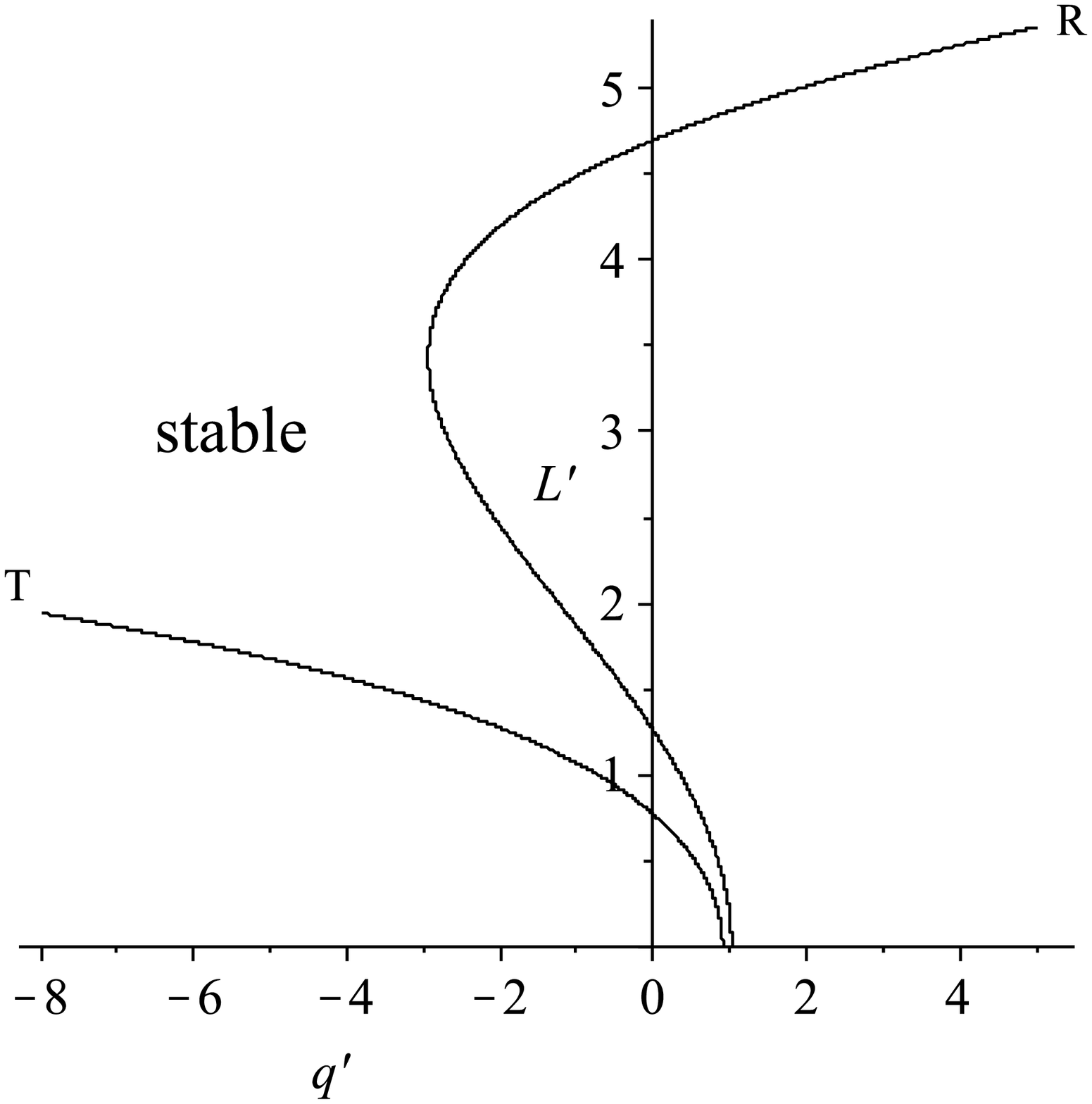}

(d)\includegraphics[width=0.25\linewidth]{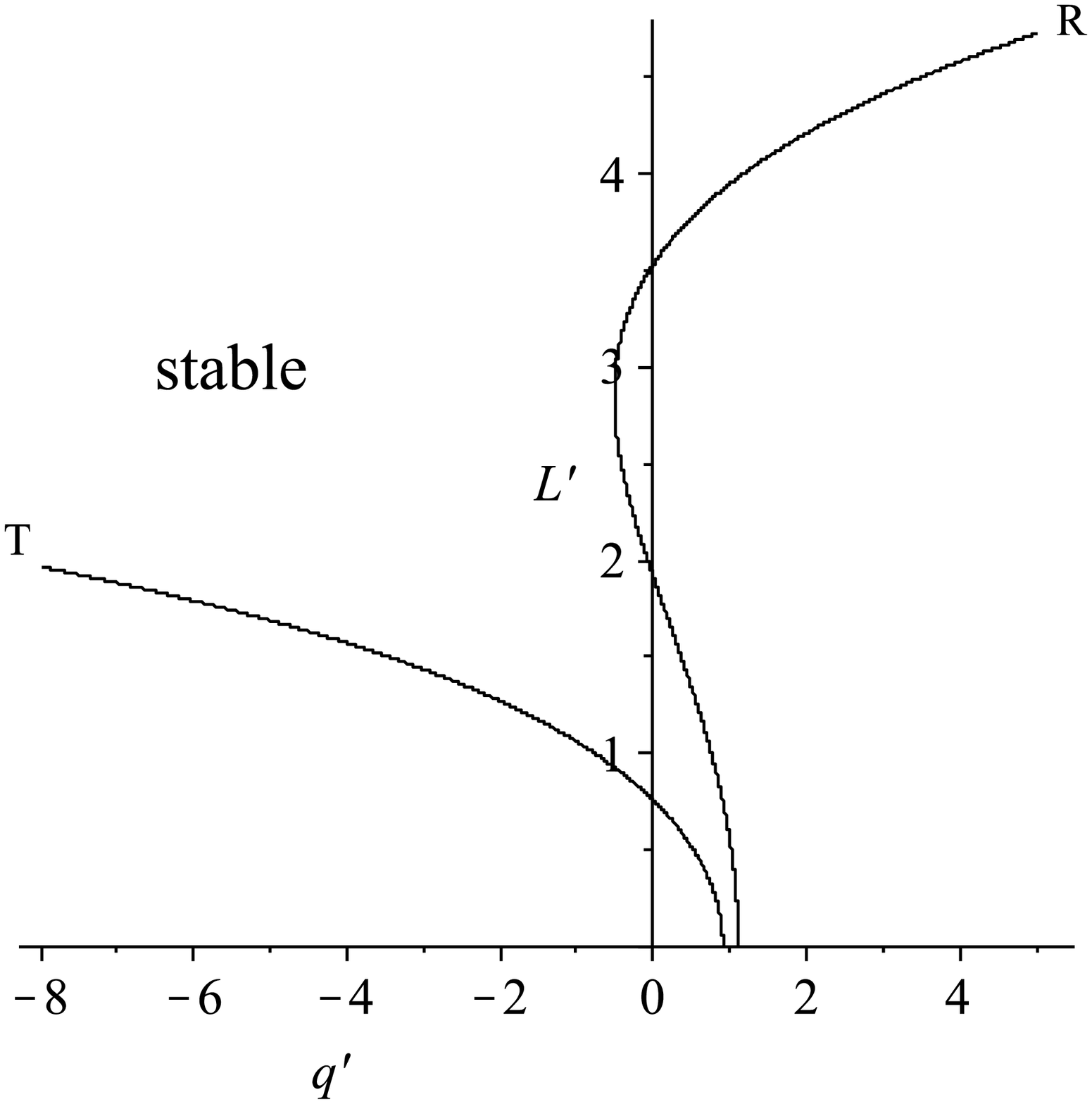}
(e)\includegraphics[width=0.25\linewidth]{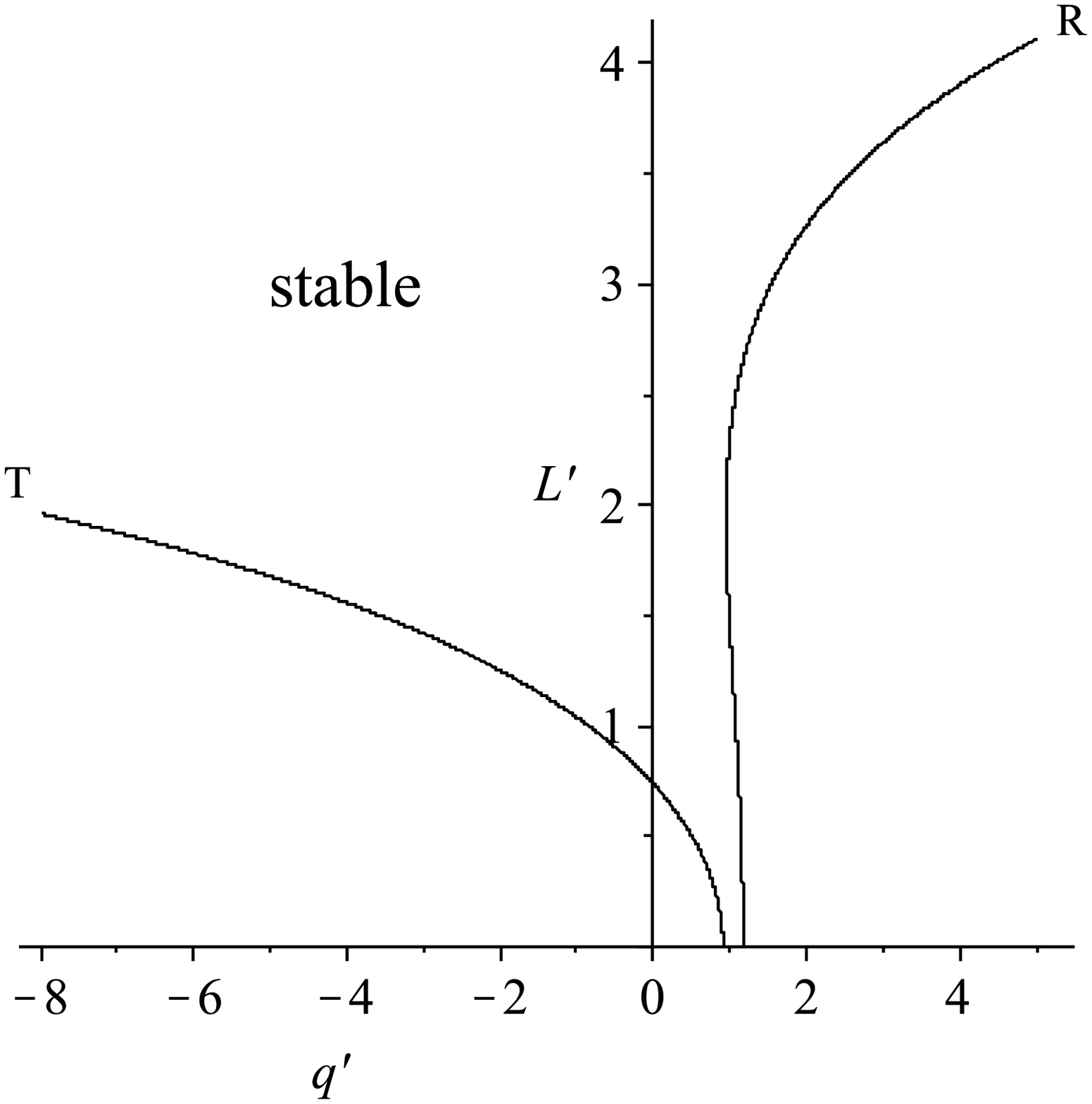}
(f)\includegraphics[width=0.25\linewidth]{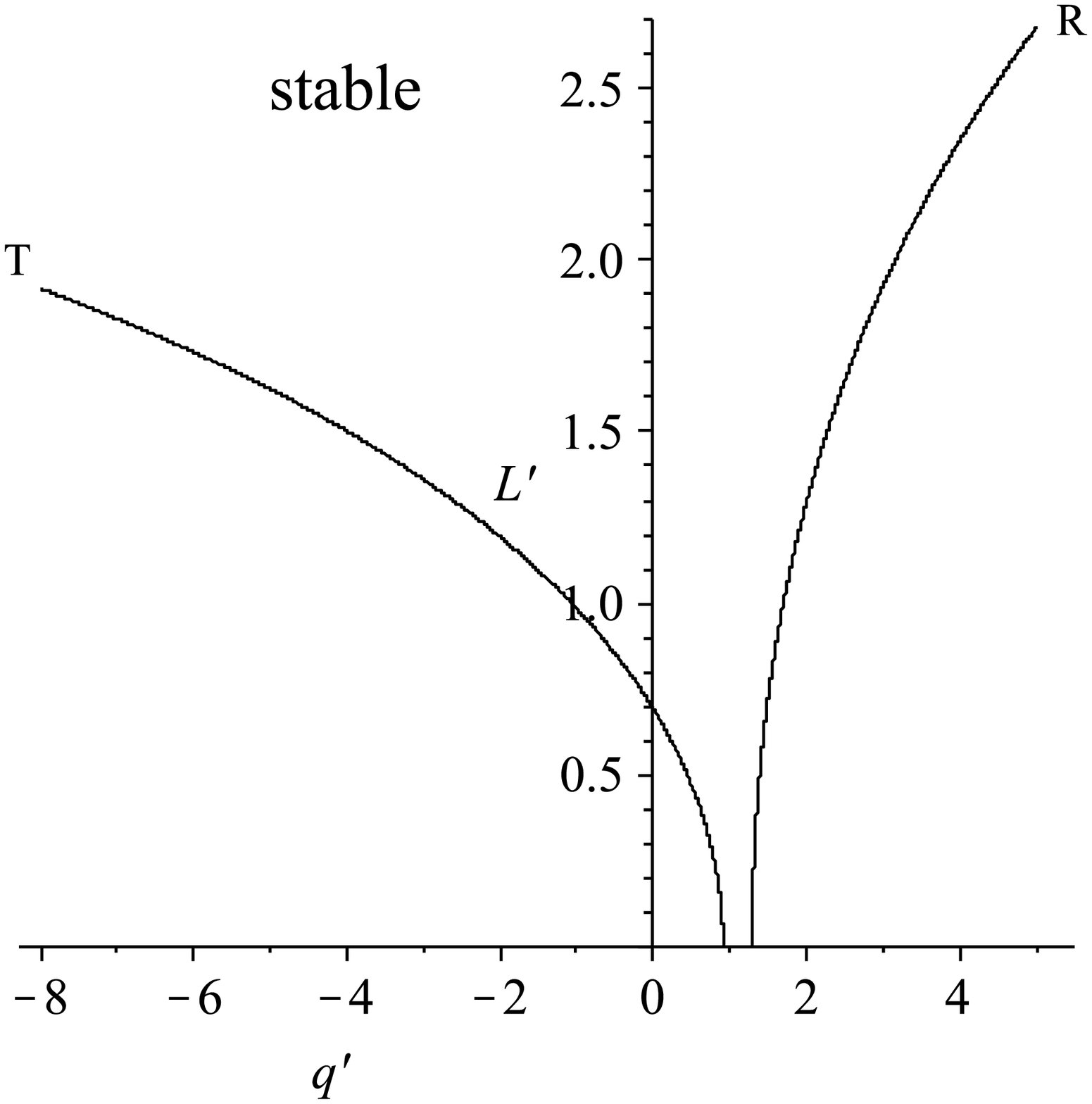}

\caption{The stability boundaries for traveling waves according to the 
amplitude equations \eqref{eq:P}--\eqref{eq:a}, in the special case 
$\beta'=0$, for different values for $\alpha'$: (a) $\alpha'=1$, (b) 
$\alpha'=3$, (c) $\alpha'=4$, (d) $\alpha'=5$, (e) $\alpha'=5.7$ and (f) 
$\alpha'=7$. See the text for more details.}

\label{fig:12}
\end{figure}

Analysis of the stability boundaries to disturbances of general $L$
is rather involved, and we do not present the details here. Furthermore,
the parameter space is large enough to preclude our making general
statements; instead we consider some illustrative special cases. To
present the conclusions most generally, it is helpful to introduce
$q'=q/r_2$, $L'=L/r_2^{1/2}$, $\alpha'=\hat{\alpha}/r_2^{1/2}$,
$\beta'=\hat{\beta}/r_2^{1/2}$ and $v'=\hat v/r_2^{1/2}$.

Let us begin by considering the special case $\beta'=0$. 
Figure~\ref{fig:12} shows where traveling waves with different values of 
$q'$ are stable and unstable to perturbations with wavenumbers $L'$; each 
panel in the figure corresponds to a different choice of $\alpha'$. In 
understanding the sequence of transitions in the topology of the various 
panels, it is helpful to first consider the behavior of the stability 
boundaries for small $\alpha'$ (and hence small $v'$), in particular in 
the $L'=0$ limit. We have seen above that the right-hand stability curve 
(labeled $R$) intersects the $q'$ axis at $q'_+=11/12+v'^2/1152$. For small 
$\hat v$, it follows from (\ref{eq:sig2}) that the left-hand stability 
curve (labeled $T$) intersects the $q'$ axis at 
$q'_-\sim91/144+5|v'|/26568^{1/2}$. Thus as $\alpha'$ is increased from 
zero, $q'_-$ moves to the right more rapidly than does $q'_+$. Eventually, 
at some sufficiently large value of $\alpha'$, $q'_-=q'_+$, and all 
traveling waves are unstable in the limit $L'=0$. On the other hand, when 
$\alpha'$ is large, $q'_-$ halts at $q'_-=131/144$. However, $q'_+$ 
continues to increase, and this results in the appearance of a small-$L'$ 
stability region. In fact, for sufficiently large $\alpha'$, some rolls 
are stable to disturbances for all $L'$. For $0\leq\alpha'<\alpha'_c$, 
where $\alpha_c'\approx5.7$, all traveling waves are unstable. For 
$\alpha'>\alpha'_c$, a stable region appears (see Fig.~\ref{fig:12}(e)). 
Subsequently, for any value of $\alpha'>\alpha_c'$ the stable region 
becomes more apparent.

This result can be compared with the numerical stability results shown
in  Fig.~\ref{fig:32}(a), where $\alpha=1/2$. The stability condition 
$\alpha' > \alpha_c'\approx5.7$  (where $\alpha' = \alpha/\sqrt{r}$) 
corresponds to $r < (\alpha/5.7)^2 = 0.0077$, showing remarkably good
agreement with the upper limit of the stable region in
Fig.~\ref{fig:32}(a).

\begin{figure}
(a)\includegraphics[width=0.25\linewidth]{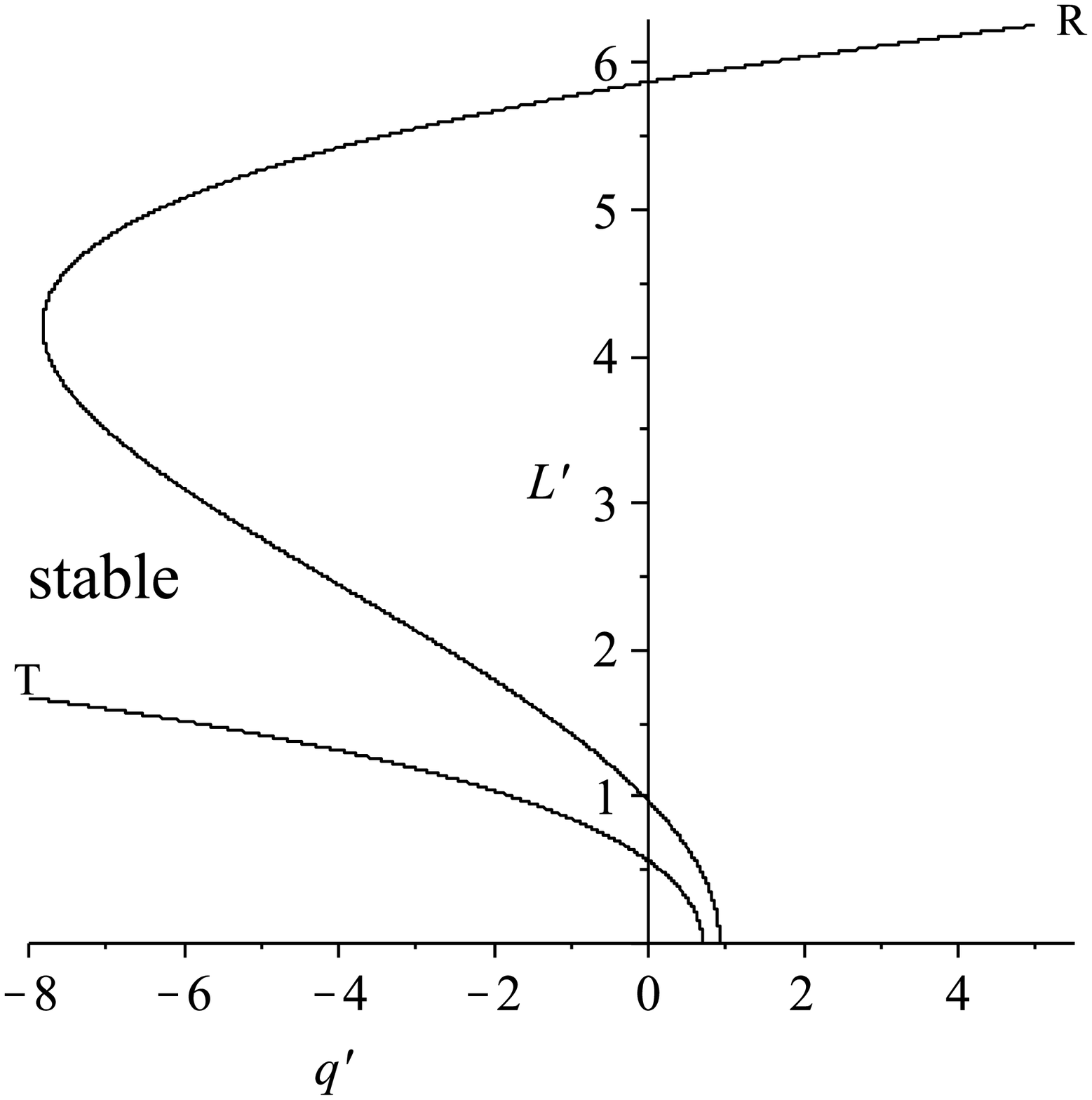}
(b)\includegraphics[width=0.25\linewidth]{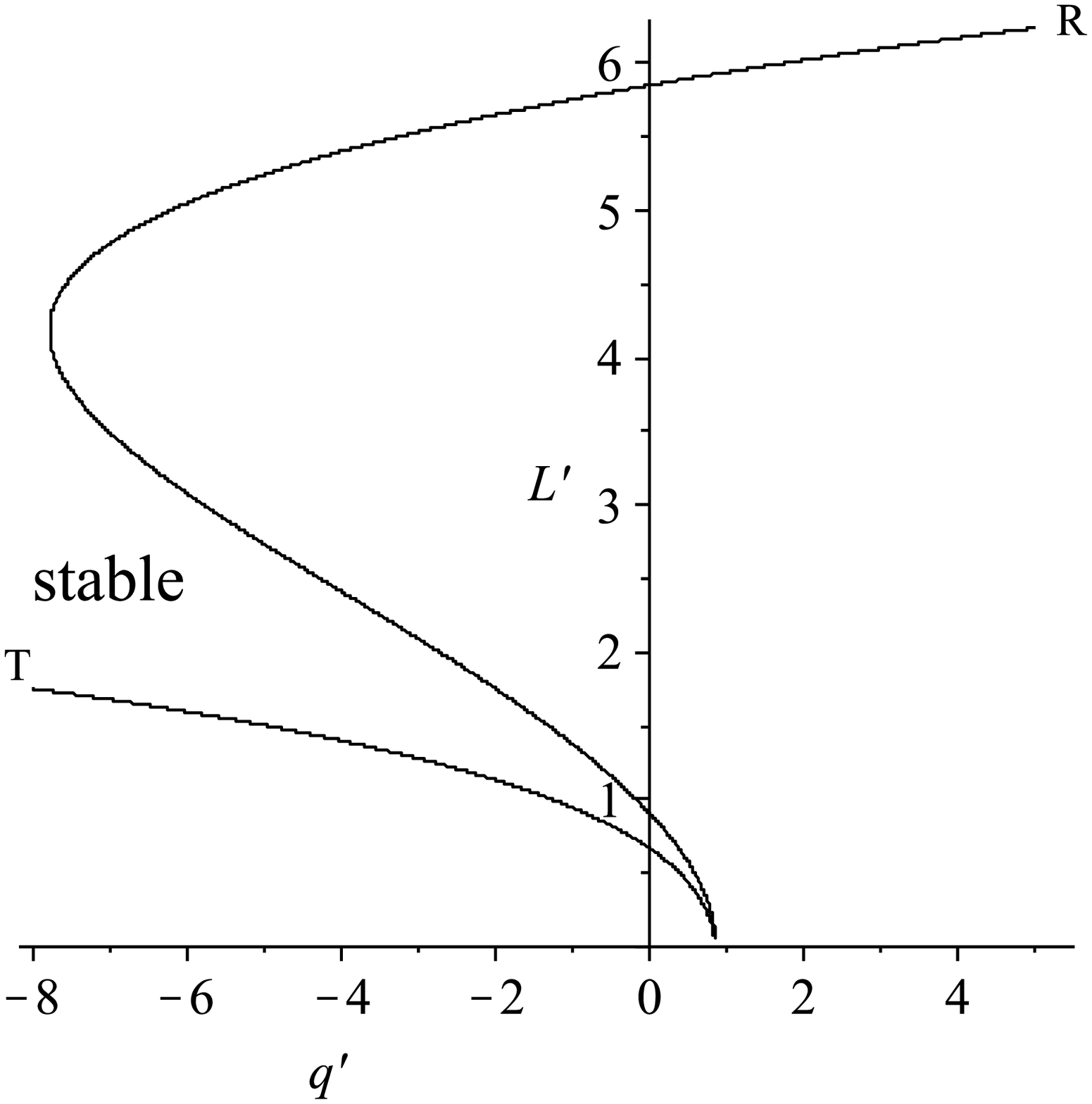}
(c)\includegraphics[width=0.25\linewidth]{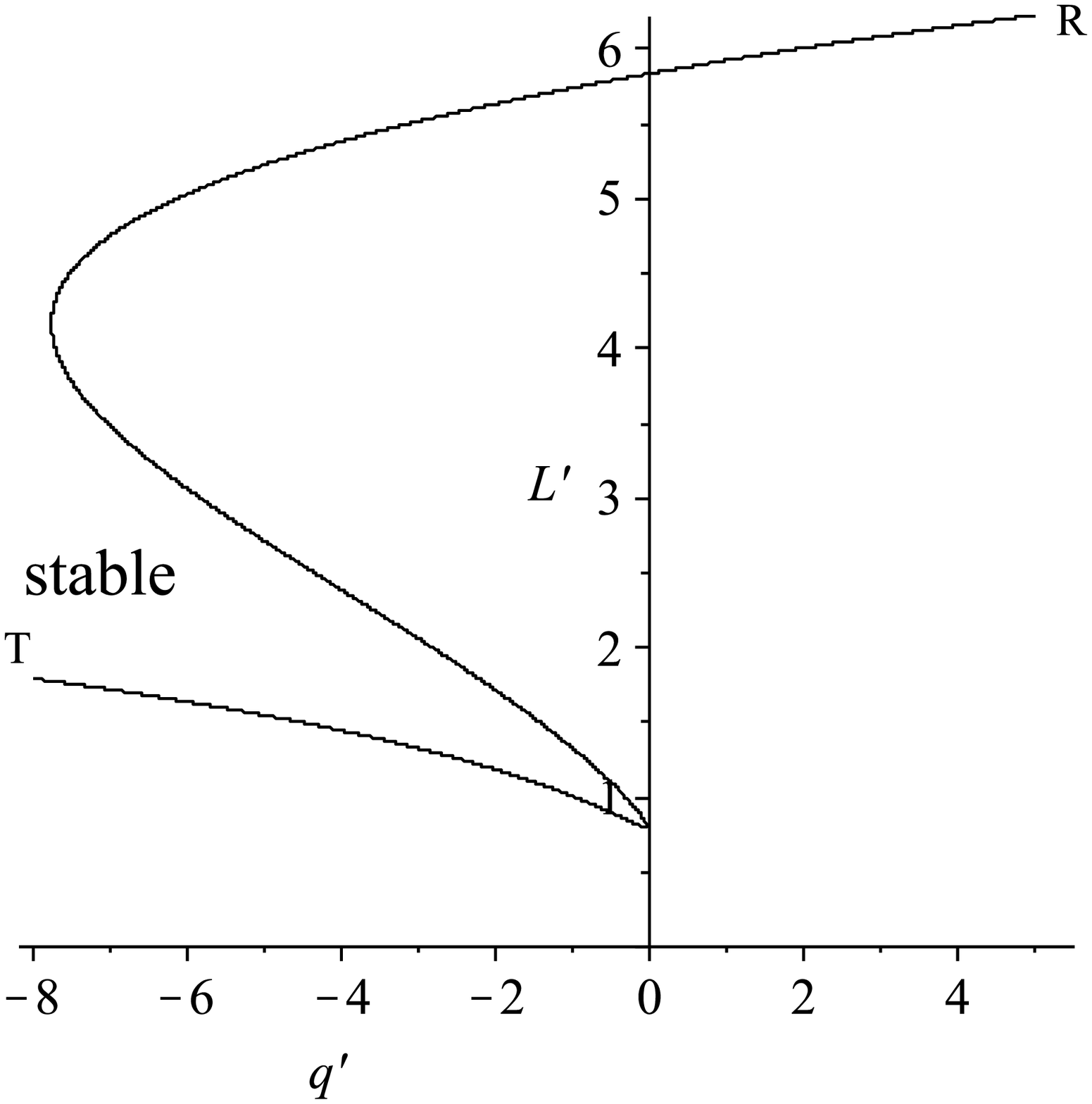}

(d)\includegraphics[width=0.25\linewidth]{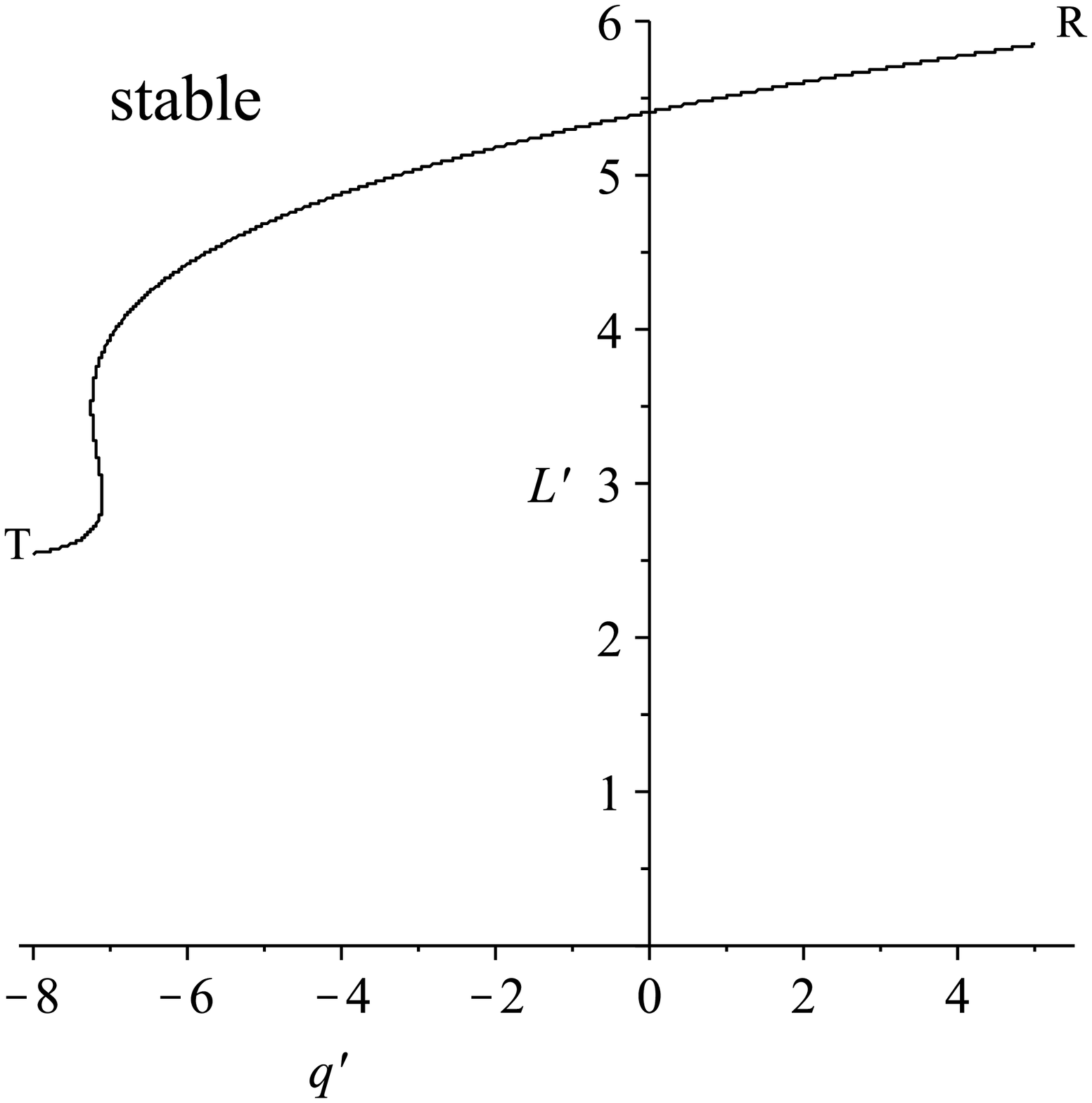}
(e)\includegraphics[width=0.25\linewidth]{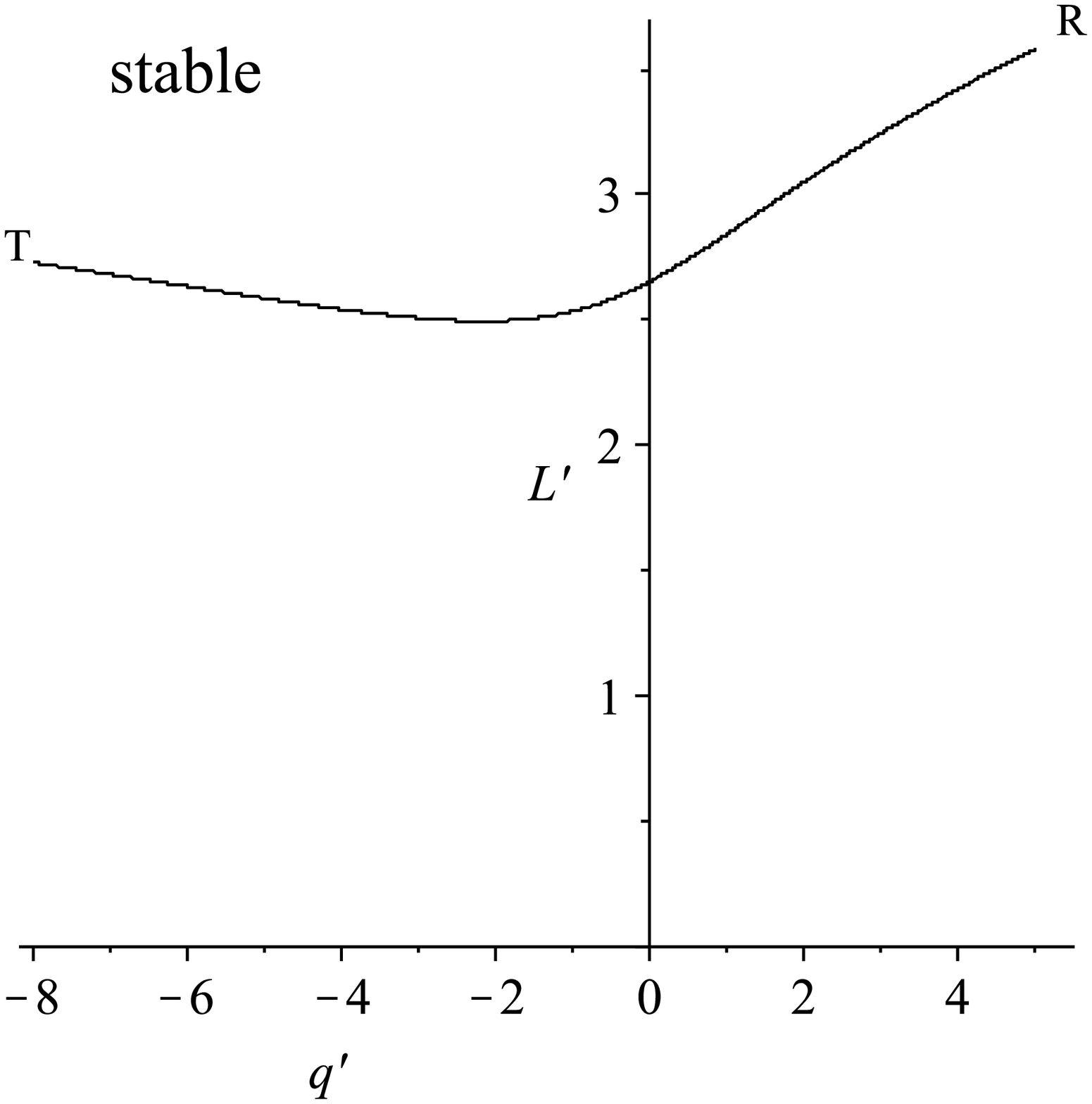}
(f)\includegraphics[width=0.25\linewidth]{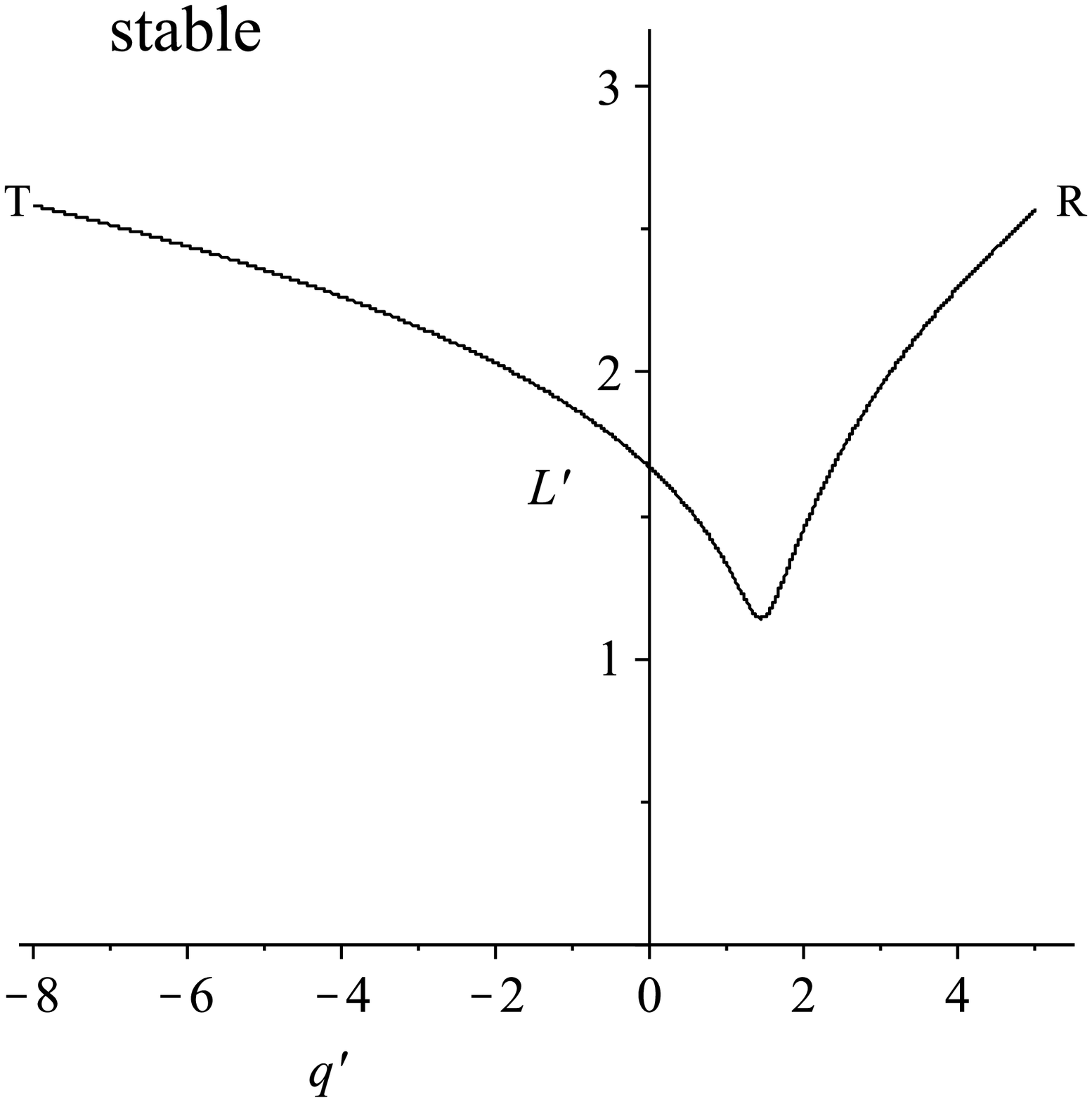}

(g)\includegraphics[width=0.25\linewidth]{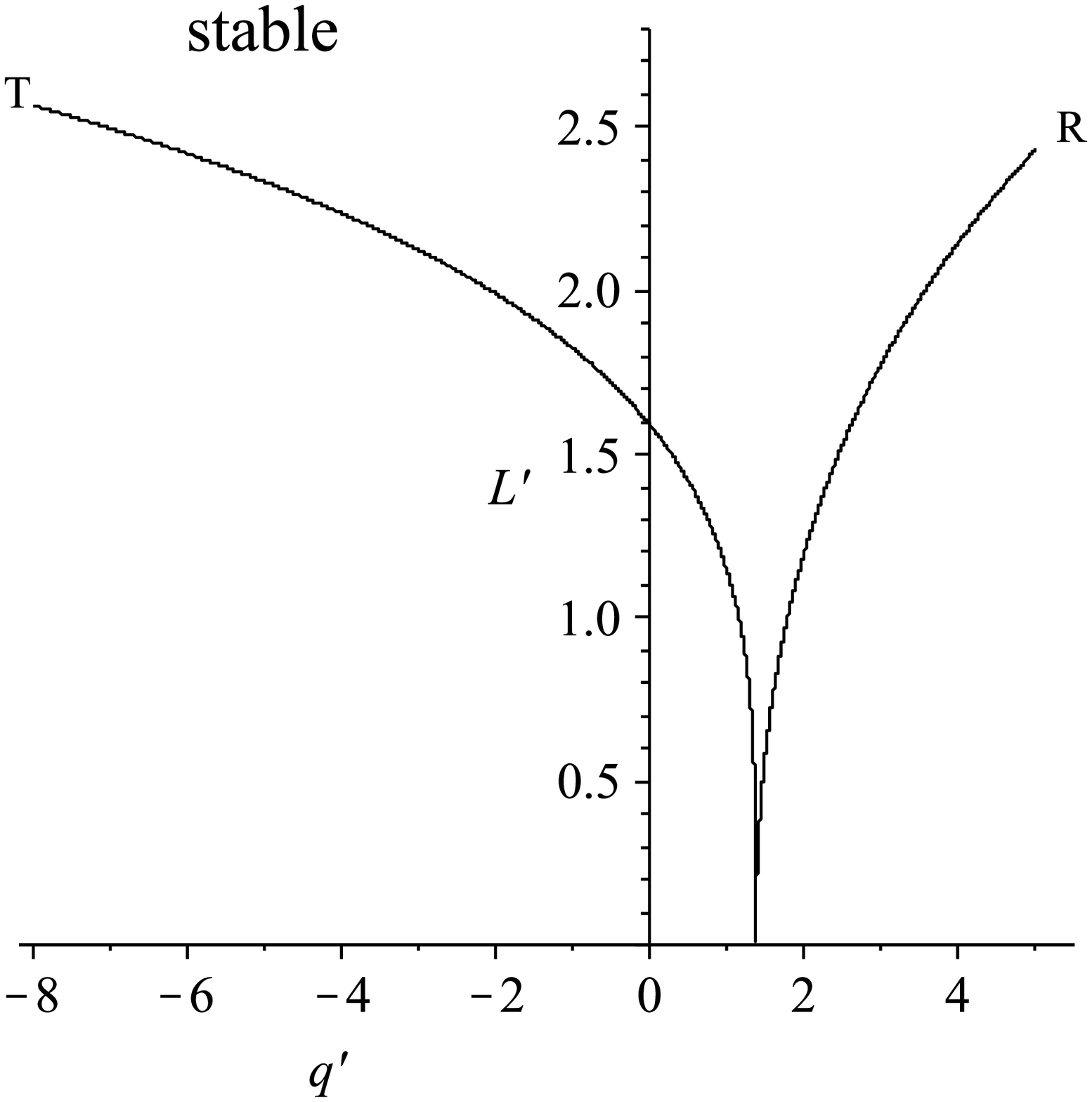}
(h)\includegraphics[width=0.25\linewidth]{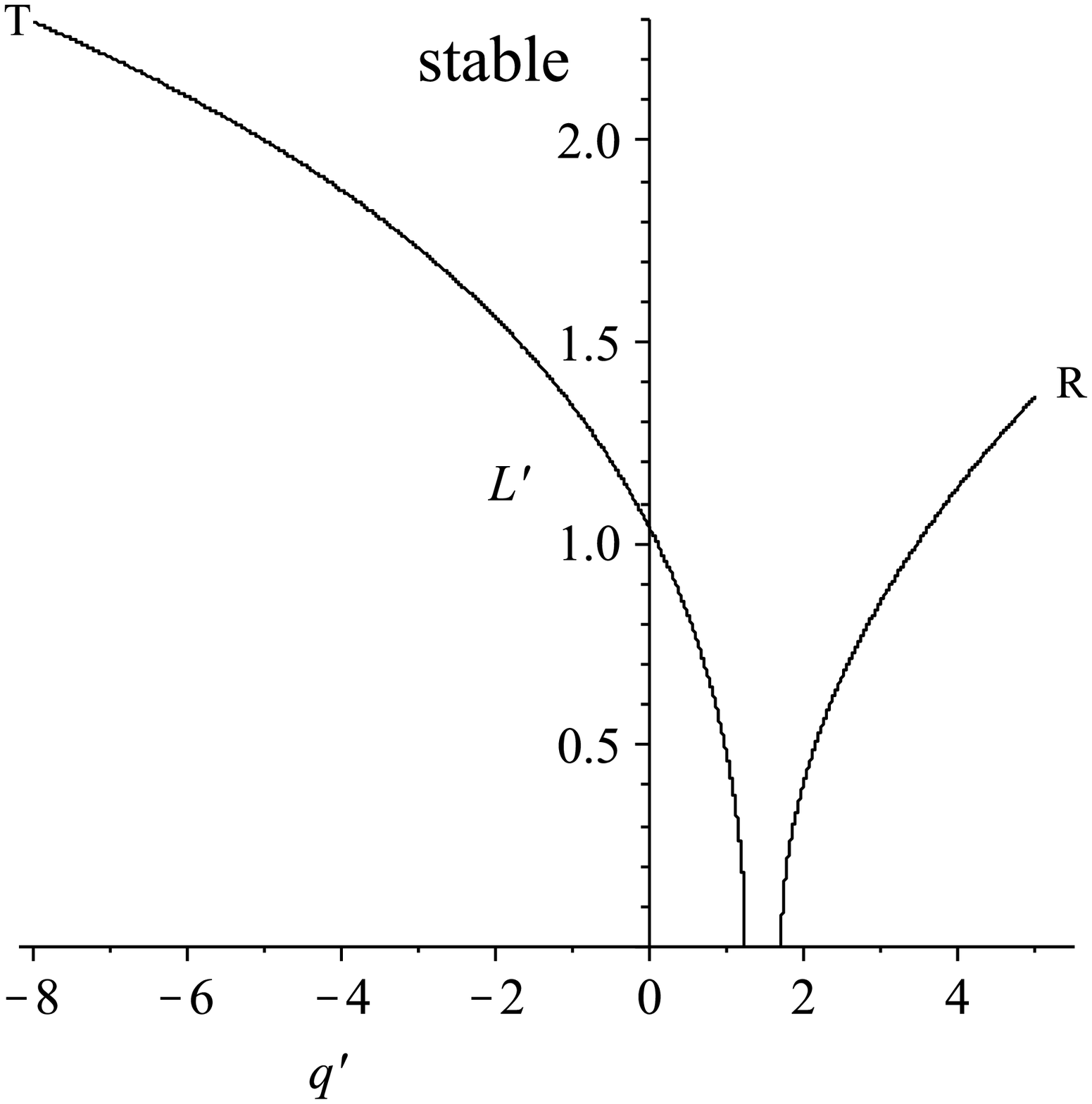}

\caption{The stability boundaries for traveling waves according to the 
amplitude equations \eqref{eq:P}--\eqref{eq:a}, in the special case 
$\alpha'=0$, for different values for $\beta'$: (a) $\beta'=0.2$, (b) 
$\beta'=0.476$, (c) $\beta'=0.6$, (d) $\beta'=2$, (e) $\beta'=4.5$, (f) 
$\beta'=5$, (g) $\beta'=5.0607$, (h) $\beta'=6$. See the text for more 
details.}

\label{fig:12b}
\end{figure}

If instead we consider the special case $\alpha'=0$, with $\beta'>0$, we
find a broadly similar picture, in that all traveling waves are unstable
when $\beta'$ is small, but some eventually stabilize, once $\beta'$ is
sufficiently large. From Fig.~\ref{fig:12b} it is apparent that the two
stability boundaries $R$ and $T$ intersect, coalesce, then lift off from
the $q'$ axis as $\beta'$ is increased. Ultimately they re-attach to the
$q'$ axis, when $\beta'=\beta_c'$, where $\beta_c'\approx5.06$ as shown in
Fig.~\ref{fig:12b}(g). For $\beta'>\beta_c'$, there is a region of
stable traveling waves.

Let us now express the results above in a form more illuminating for 
comparison with our earlier numerical secondary stability calculations 
(Sec.~\ref{sec:secstab}). As an example, we set $\hat{\alpha}=1$ and 
$\hat{\beta}=0$ and consider the limit of small $L$, looking for regions 
of stable waves as $r_2$ is varied.

\begin{figure}

(a)
\includegraphics[width=0.3\linewidth]{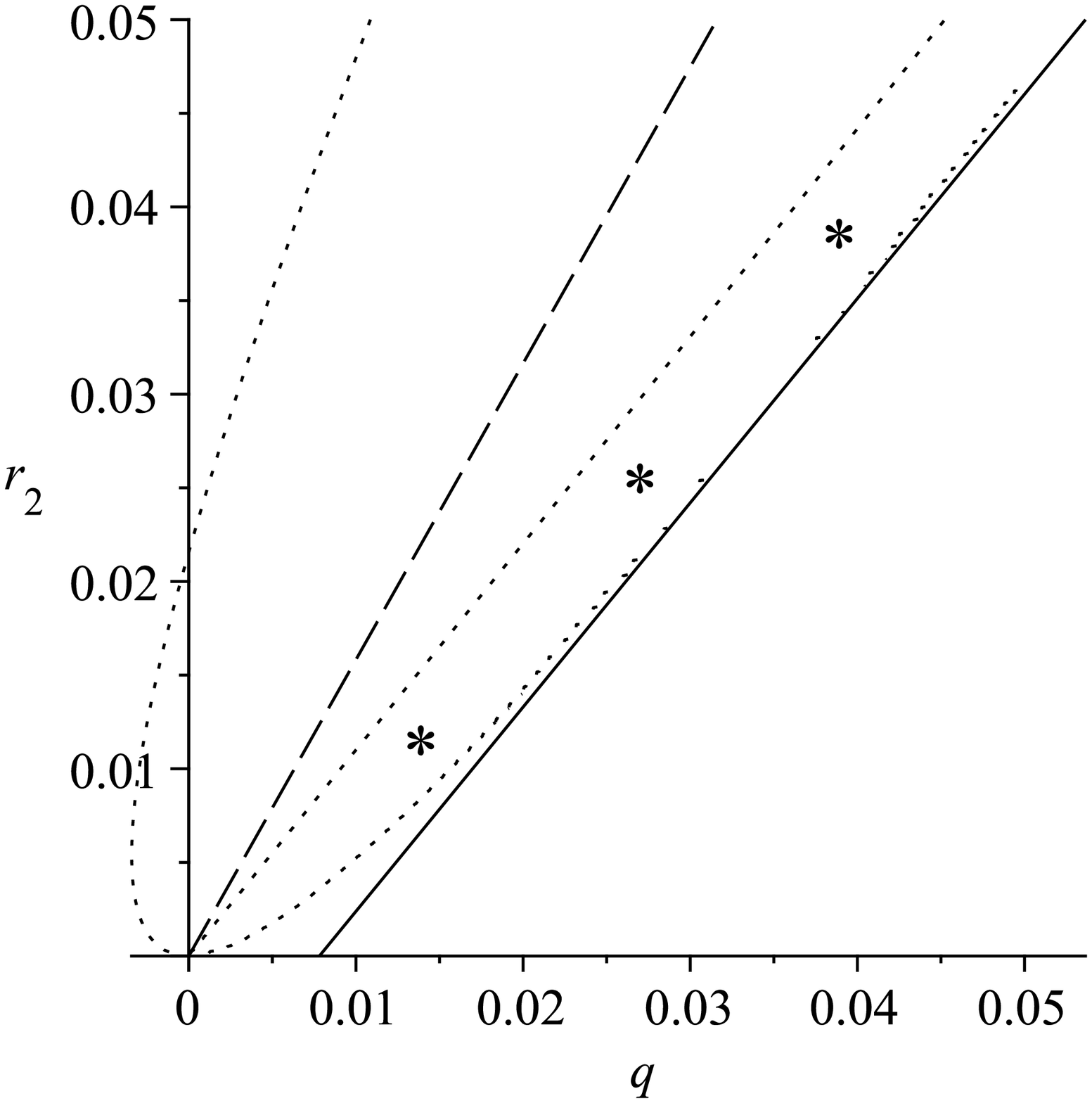}
(b)
\includegraphics[width=0.3\linewidth]{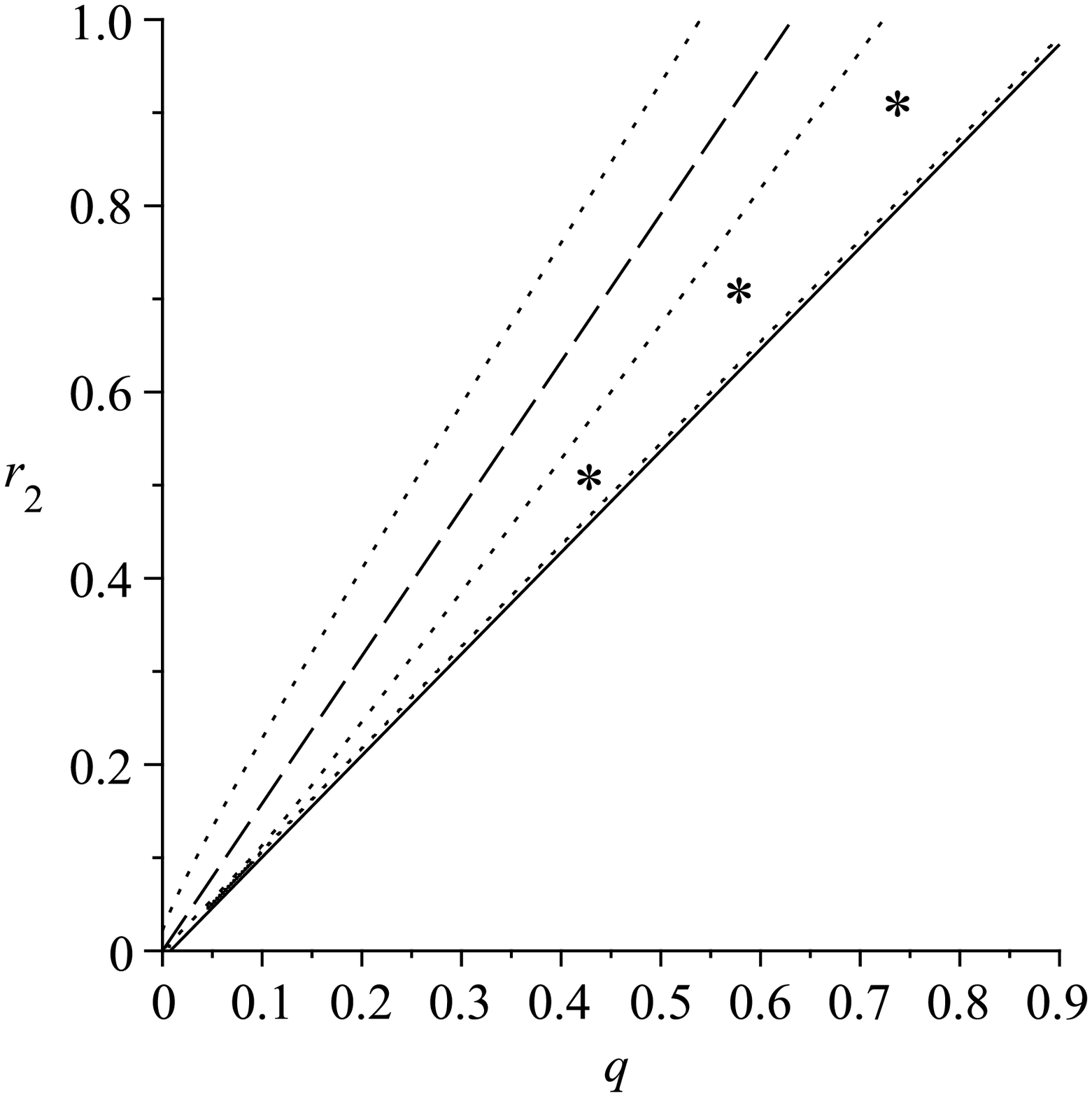}

\caption{The region of secondary stability of traveling waves is marked 
with asterisks; for details refer to the text. The solid line shows where 
$\sigma_1$ is purely imaginary; to the right of this line, the traveling 
waves are certainly unstable. The dashed line shows where $q=91 r_2/144$; 
to the left of this line, traveling waves are also certainly unstable. The 
dotted lines show where $\sigma_2=0$.}

\label{fig:15}
\end{figure}

From (\ref{eq:sig1}), rolls are unstable as long as $q>11r_2/12+\hat 
v^2/1152$. If $q<11r_2/12+\hat v^2/1152$, then $\sigma_1$ is purely 
imaginary and hence $\sigma_2$ must be considered. From (\ref{eq:sig2}) we 
have definite instability if $r_2>144q/91$. In addition to these rather 
blunt conditions, the sign of $\sigma_2$ must also be considered in order 
to determine the stable region. Figure~\ref{fig:15} shows the curves
$q=11r_2/12+\hat v^2/1152$ (solid line), $q=91r_2/144$ (dashed line) and 
$\sigma_2=0$ (dotted lines). Any region of stability must lie between the 
solid and dashed lines. After checking carefully the signs of the 
eigenvalues, we find that the stable region (indicated by the asterisks in 
the figure) lies between the two dotted lines in the upper and lower parts 
of the graph, and between the dotted and solid lines for a small range 
of intermediate values of $r_2$ (see Fig.~\ref{fig:15}). Although they 
appear almost parallel in Fig.~\ref{fig:15}(a), for large $r_2$, as in 
Fig.~\ref{fig:15}(b), the two sides of the secondary stability region are 
no longer approximately parallel.

Note the qualitative similarity between the shapes of the stable
regions in Fig.~\ref{fig:15} and Fig.~\ref{fig:32}.

The question remains of whether or not this stable region extends to 
indefinitely large values of $r_2$. To investigate the large-$r_2$ 
behavior of the stability region, we consider large $r_2$ with $q=O(r_2)$, 
motivated by the observation, from Fig.~\ref{fig:15}(b), that stable 
rolls lie in some region between straight lines in $(q,r_2)$ parameter 
space. In this limit the stability condition from (\ref{eq:sig1}) 
simplifies to $q < 11 r_2/12$, while $\sigma_2 = 91/2-72 q/r_2 + 
O(r_2^{-1/2})$. Hence we can conclude that the region of stable waves for 
small $L$ and large $r_2$ is
 \beq
91r_2/144 < q < 11r_2/12.
 \eeq
In summary, the results of this section show that when  $\alpha$
and $\beta$ are $O(\ep)$, there can be a narrow region of stable
traveling waves near $k=1$, and that there is no upper limit on the size of
$r_2$ allowing  stable rolls.  

For even smaller values of $\alpha$ and $\beta$, of order $\ep^2$, 
we have checked that  $\alpha$ and $\beta$ do not appear in the
leading order amplitude equations, so in that case all traveling
waves are unstable, as in the non-dispersive case.

\subsection{Traveling waves close to the marginal curve}

We now turn to the second case in which $a_0^2q$ may be small: in the 
region close to the marginal stability curve. Following an analysis similar
to that for the dissipative Nikolaevskiy equation~\cite{CMdamp}, we find
that, in contrast to the dissipative case (in which a narrow region of
stable rolls exists close to the marginal curve~\cite{CMdamp}), here all
traveling waves are unstable near the marginal curve.

\section{Numerical simulations of the dispersive Nikolaevskiy equation}
\label{sec:numsim}

\begin{figure}
(a)\includegraphics[width=0.25\linewidth]{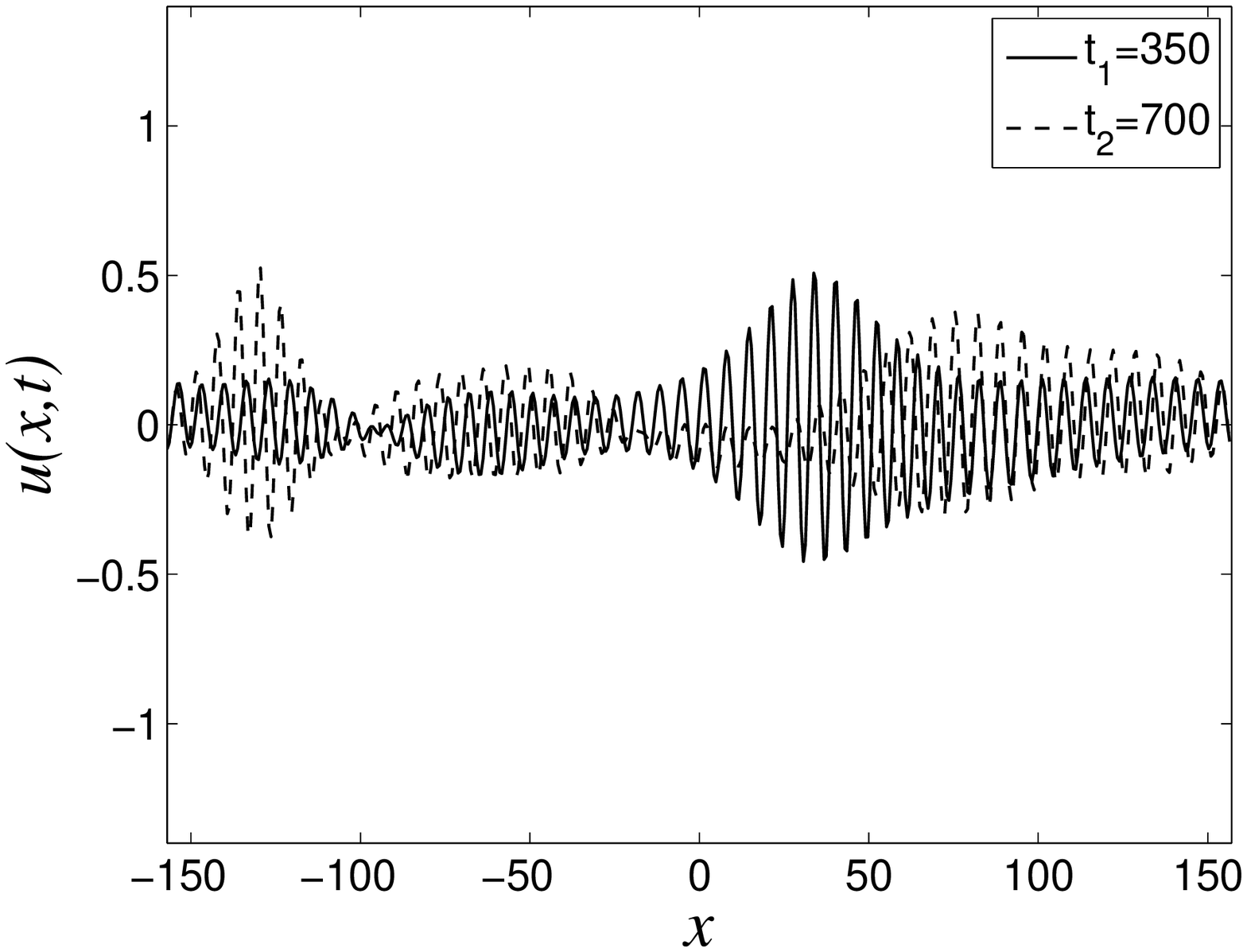}
(b)\includegraphics[width=0.25\linewidth]{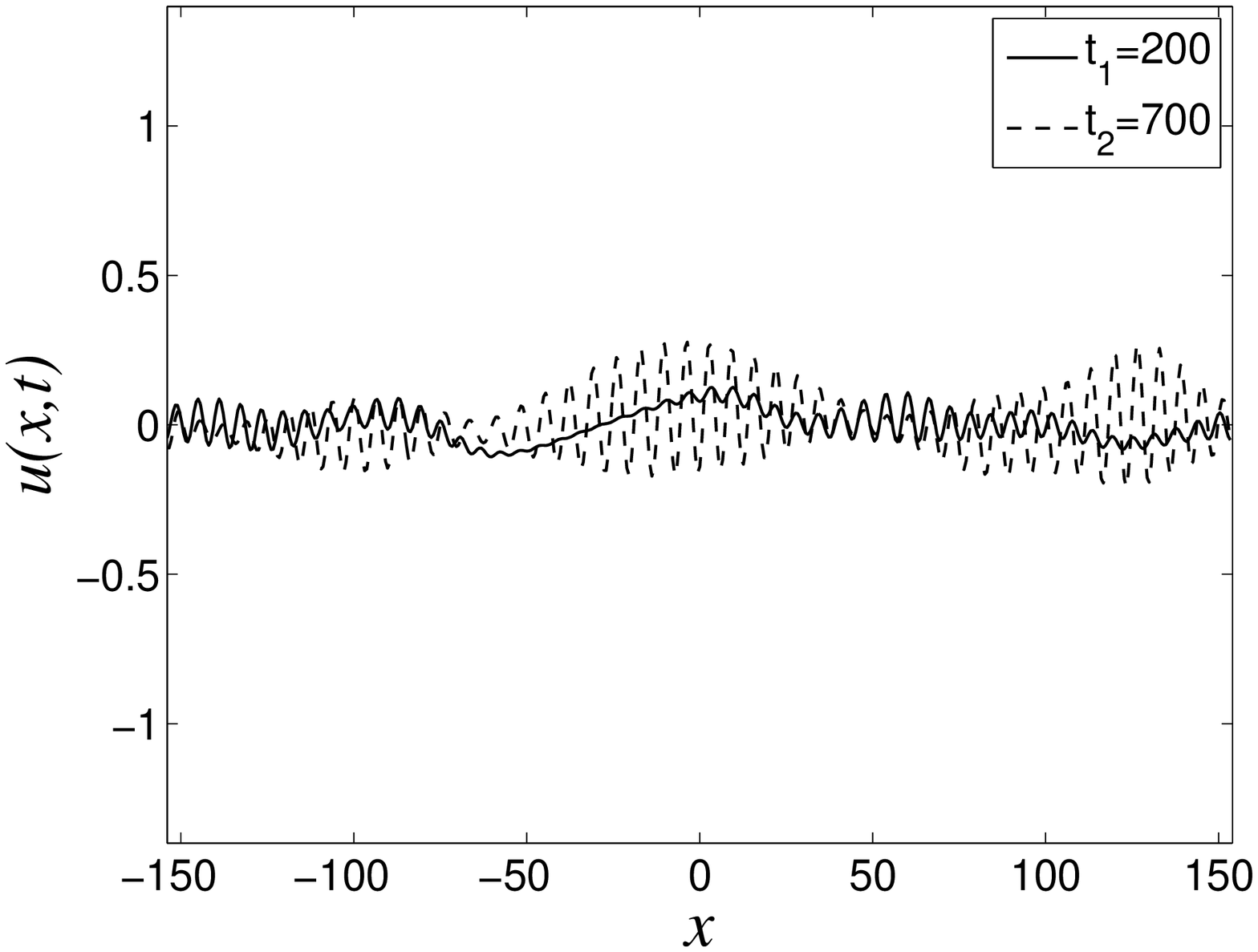}
(c)\includegraphics[width=0.25\linewidth]{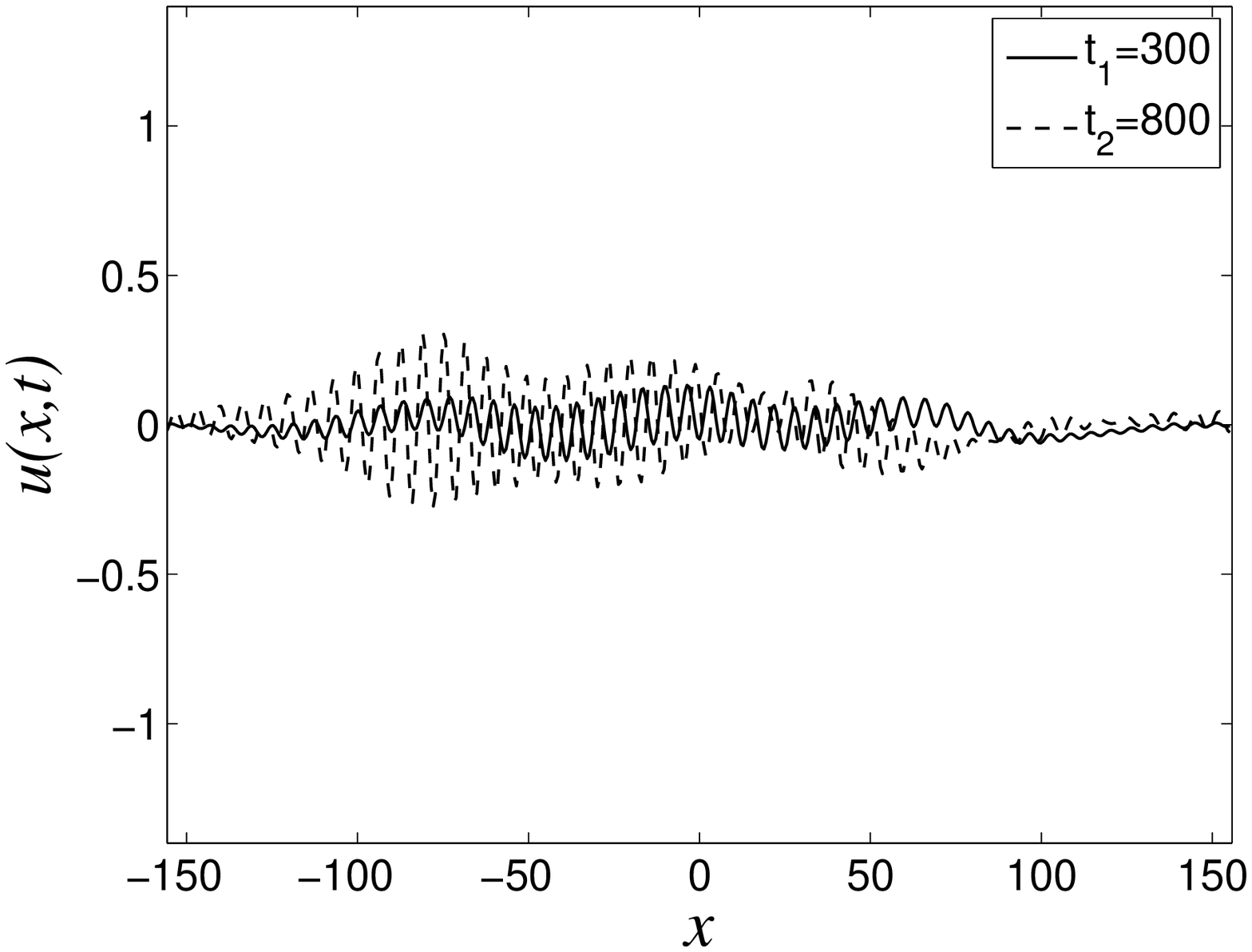}

(d)\includegraphics[width=0.25\linewidth]{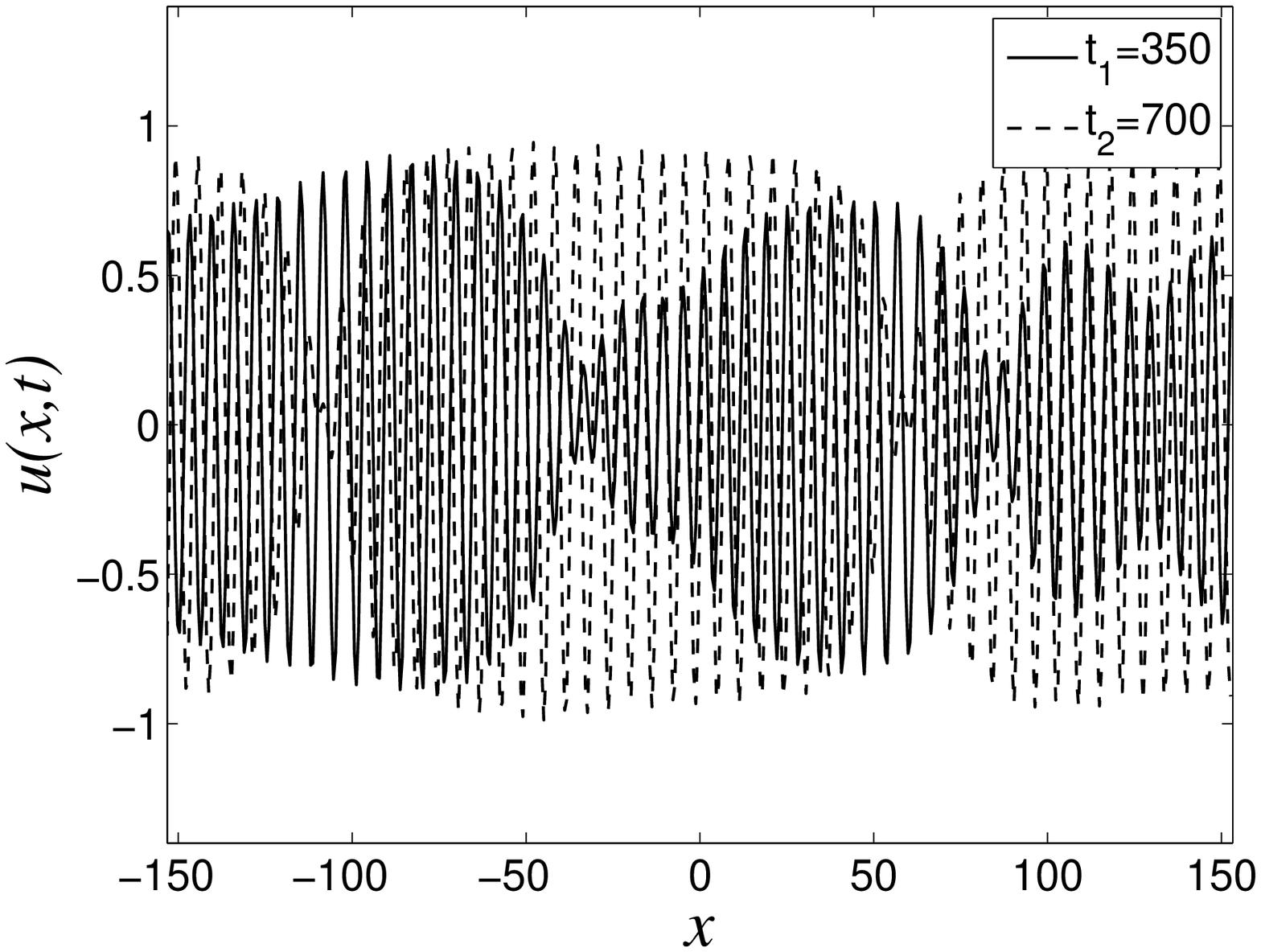}
(e)\includegraphics[width=0.25\linewidth]{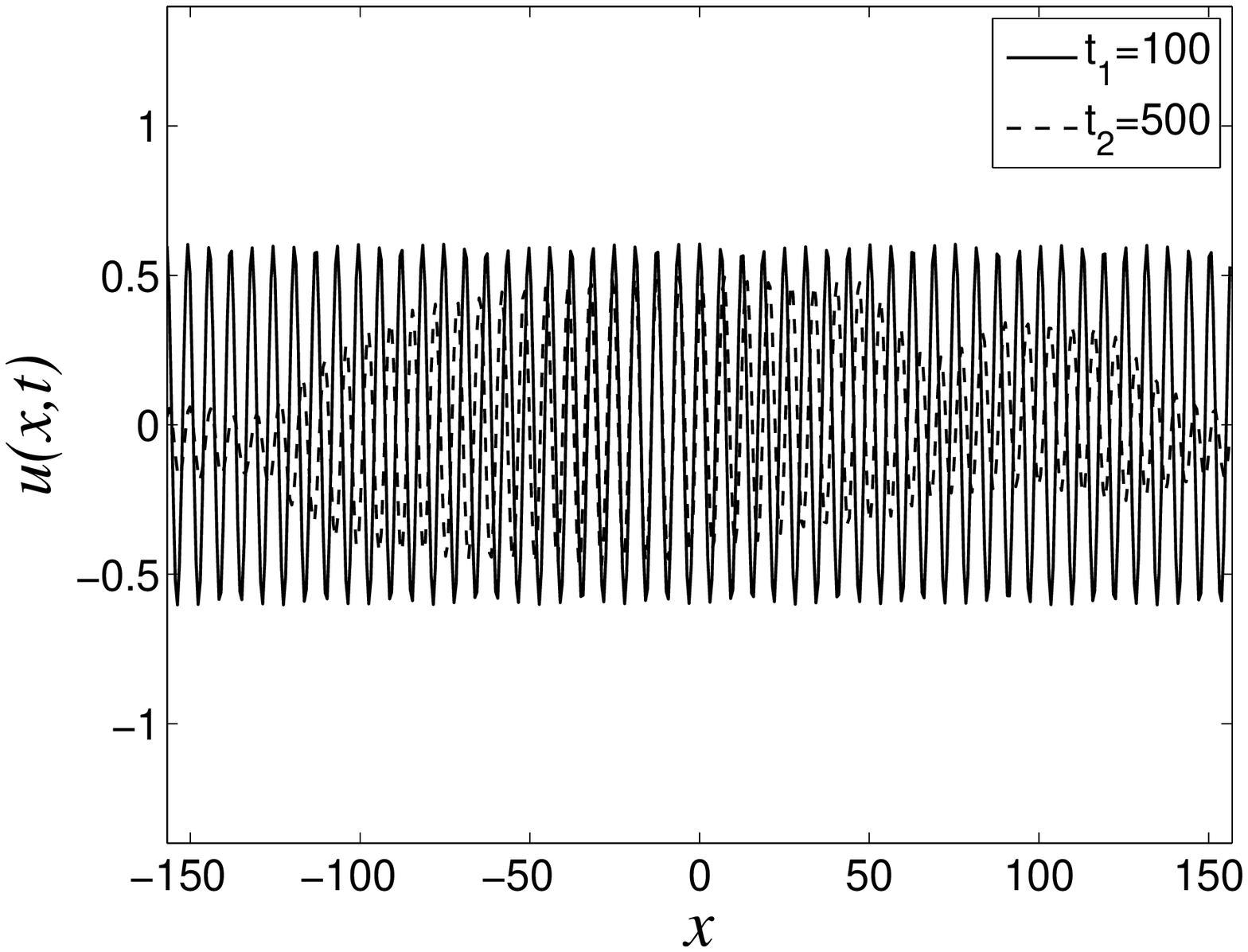}

\caption{Snapshots of the numerical solutions of \eqref{eq:nikd}.
Parameter values are: (a) $\alpha=2$, $\beta=1$, $r=0.01$ and $k=1$;
(b) $\alpha=0.3557$, $\beta=0.1778$, $r=0.01$ and $k=1.02$;
(c) $\alpha=0.2$, $\beta=0$, $r=0.01$ and $k=1.0087$;
(d) $\alpha=0$, $\beta=0.5$, $r=0.01$ and $k=1.025$;
(e) $\alpha=0.25$, $\beta=0$, $r=0.0025$ and $k=1.00125$.
The times of the snapshots are indicated in the insets.}
\label{fig:13}
\end{figure}

To illustrate some of the consequences of the results of the preceding 
sections, we have carried out numerical simulations of the dispersive 
Nikolaevskiy equation, using a pseudospectral method, with exponential 
time stepping~\cite{CoxMatt}, of which a small sample are presented here. 
The initial condition is taken to be a traveling wave with a given 
wavenumber $k$ (approximated as a cosine of the amplitude given by 
(\ref{eq:TWamp})), plus small random noise, and the domain size is 
$D=100\pi/k$. Figure~\ref{fig:13} illustrates in order strong (a), 
intermediate (b) and weak dispersion (c)--(e). The values of $\alpha$, 
$\beta$ and $k$ are chosen in each case to correspond to traveling waves 
that are predicted to be unstable by the asymptotic analysis.

Figure~\ref{fig:13} (a) shows the case of strong dispersion, with 
$\alpha=2$ and $\beta=1$, and wave number $k=1$ and $r=0.01$. The 
stability analysis of \eqref{eq:strAcanon} predicts that the rolls are 
unstable, since $\alpha=2$ and $\beta=1$ lie in the unstable region in 
Fig.~\ref{fig:+-}; the numerical simulation agrees with this asymptotic 
result.

In Fig.~\ref{fig:13}~(b) an example of intermediate dispersion is 
simulated, where $\alpha=2\epsilon^{3/4}$ and $\beta=\epsilon^{3/4}$, 
$r=\epsilon^2$ and $k=1+\epsilon q$, for $q=0.2$ and $\epsilon=0.1$. It is 
known from the asymptotic results of Sec.~\ref{sec:int} that $\alpha$ and 
$\beta$ being $O(\epsilon^{3/4})$ with wave number $k=1+\epsilon q$ will 
result in unstable traveling wave solutions, which agrees with the 
simulation shown in Fig.~\ref{fig:13}~(b).

To show the effects of weak dispersion with wave number $k=1+\epsilon^2q$ 
we take $r=0.01$ and $\epsilon=0.1$. Figure~\ref{fig:13} (c) shows the 
case $\alpha=2\epsilon$, $\beta=0$, $q=0.87$, while in Fig.~\ref{fig:13} 
(d) the parameter values are $\alpha=0$, $\beta=5\epsilon$, $q=2.5$. Rolls 
should in each case be unstable, according to the analysis of 
Sec.~\ref{sec:3}, and this is confirmed by the numerical simulations.

Figure~\ref{fig:13}~(e) represents weak dispersion, with $\alpha=\epsilon$ 
and $\beta=0$. The wavenumber is $k=1+\epsilon^2q$ and $r=\epsilon^2r_2$, 
for $\epsilon=0.25$, $q=0.02$ with $r_2=0.04$. These values of $r_2$ and 
$q$ lie in the unstable region given in Fig.~\ref{fig:15}, and the 
simulations support this prediction of instability.

\section{Conclusions}

We have examined the stability of spatially periodic solutions to the 
dispersive Nikolaevskiy equation, which is the original model introduced 
by Nikolaevskiy~\cite{Nik89} for seismic waves. The reincorporation of 
dispersive effects stands in contrast to most studies subsequent to 
Nikolaevskiy's paper. We have shown how the instability of {\em all} 
spatially periodic solutions at the onset of pattern formation in the 
more-often treated, nondispersive version is modified by the presence of 
dispersive terms. Our results have been achieved through both a numerical 
calculation of the secondary stability boundary for the traveling wave 
solutions and an asymptotic treatment of three particular scalings in 
$\ep$ for the dispersive terms. The secondary stability diagrams (``Busse 
balloons'') can be rather complicated, and can depend sensitively on the 
size of the dispersive terms.

Our consideration of the case $\alpha,\beta=O(1)$ can be interpreted as
giving information about the bottom of the secondary stability diagram
obtained in $(k,r)$ parameter space for fixed $\alpha$ and $\beta$. Two
cases were found: either all traveling waves are unstable at the bottom
of the diagram, or there is a symmetrical, Eckhaus-like region of stable
traveling waves, right down to onset at $r=0$ (although the width of the
region of stable rolls does not stand in the usual Eckhaus ratio to the
width of the existence region of rolls).

The separate analysis for smaller values of $\alpha,\beta$ can be
interpreted as shedding light on the upper parts of the fixed-$\alpha$,$\beta$
stability diagram in $k,r$ parameter space. We have shown that for
small $\alpha,\beta$, a narrow region of stable waves may exist near
$k=1$.  However, beyond the range of validity of the asymptotic
analysis, the numerical stability results show the
complicated nature of the secondary stability boundaries, so we are
unable to draw any significant general conclusions about the form
of the secondary stability diagram, limiting ourselves to some specific
examples. Things are
further complicated by the fact that rolls predicted to be stable by the
asymptotics may in fact turn out to be unstable  when the full numerical
calculation is performed, since the asymptotics concerns only long-wavelength
instabilities, and other, short-wavelength instabilities may turn out to
be present.

In this paper, we have said little about the behavior of time-dependent
solutions of the dispersive Nikolaevskiy equation. However, it appears
from our numerical simulations that when all waves are
unstable, chaotic states are found that have a similar behavior to
that found in the non-dispersive Nikolaevskiy
equation~\cite{MattCox00,SakTan,TribTsu}.




\end{document}